\documentclass[runningheads]{llncs}
\usepackage{graphics,color}
\usepackage{graphicx}
\usepackage{amssymb,amsmath,amsfonts,amssymb}
\usepackage{amsmath,amssymb,mathrsfs,bm,url,times}
\usepackage{subfigure}
\usepackage{multirow} 
\newtheorem{phenomenon}{Phenomenon}
\newtheorem{insight}{Insight}

\begin{document}

\title{A Characterization of Cybersecurity Posture from Network Telescope Data}

\titlerunning{A Characterization of Cybersecurity Posture from Network Telescope Data}

\author{Zhenxin Zhan\inst{1} \and Maochao Xu\inst{2} \and Shouhuai Xu\inst{1}}

\authorrunning{Z. Zhan and M. Xu and S. Xu}

\institute{Department of Computer Science, University of Texas at San Antonio\\ 
\email{jankins.ics@gmail.com}, \email{shxu@cs.utsa.edu} \\
\and
Department of Mathematics, Illinois State University \\
\email{mxu2@ilstu.edu}
}

\maketitle

\begin{abstract}
Data-driven understanding of cybersecurity posture is an important problem that has not been adequately explored.
In this paper, we analyze some real data collected by CAIDA's network telescope during the month of March 2013.
We propose to formalize the concept of cybersecurity posture from the perspectives of three kinds of time series:
the number of victims (i.e., telescope IP addresses that are attacked), the number of attackers that are observed by the telescope,
and the number of attacks that are observed by the telescope.
Characterizing cybersecurity posture therefore becomes investigating the phenomena and statistical properties exhibited by these time series,
and explaining their cybersecurity meanings.
For example, we propose the concept of {\em sweep-time},
and show that sweep-time should be modeled by stochastic process, rather than random variable.
We report that the number of attackers (and attacks) from a certain country dominates the total number of attackers (and attacks)
that are observed by the telescope.
We also show that substantially smaller network telescopes might not be as useful as a large telescope.

{\bf keywords:} Cybersecurity data analytics, cybersecurity posture, network telescope, network blackhole, darknet, cyber attack sweep-time, time series data
\end{abstract}

\section{Introduction}

Network telescope \cite{moore:2004:network} (aka blackhole \cite{cooke:2004:blackhole,bailey:2005:blackhole}, darknet \cite{Bailey:2006:PDM}, or
network sink \cite{yegneswaran:2004:design}, possibly with some variations) is
a useful instrument for monitoring unused, routeable IP address space.
Since there are no legitimate services associated to these unused IP addresses,
traffic targeting them is often caused by attacks.
This allows researchers to use telescope-collected data (together with other kinds of data) to study, for example,
worm propagation \cite{moore:2002:code-red,moore:2003:slammer,Shannon:2004:SWW:1018027.1018275,bailey:2005:blaster},
denial-of-service (DOS) attacks \cite{Moore:2006:IID,Hussain:2003:FCD}, and stealth botnet scan \cite{Dainotti:2012:ASS:2398776.2398778}.
Despite that telescope data can contain unsolicited --- but not necessarily malicious --- traffic that can be caused by misconfigurations
or by Internet background radiation \cite{Pang:2004:CIB,Glatz:2012:CIO,Wustrow:2010:IBR},
analyzing telescope data can lead to better understanding of {\em cybersecurity posture},
an important problem that has yet to be investigated.

\smallskip

\noindent{\bf Our Contributions.}
In this paper, we empirically characterize cybersecurity posture
based on a dataset collected by CAIDA's /8 network telescope (i.e., $2^{24}$ IP addresses) during the month of March 2013.
We make the following contributions.
{\bf First}, we propose to characterize cybersecurity posture by considering three time series:
the number of victims, the number of attackers, and the number of attacks.
To the best of our knowledge, this is the first formal definition of cybersecurity posture.
{\bf Second}, we define the notion of {\em sweep-time}, namely the time it takes for most
telescope IP addresses to be attacked at least once.
We find that sweep-time cannot be described by a probabilistic distribution, despite that a proper subset of the large sweep-times follows
the power-law distribution. We show that an appropriate stochastic process can instead describe the sweep-time.
This means that when incorporating sweep-time in theoretical cybersecurity models, it cannot always be treated as a
random variable and may need to be treated as a stochastic process.
{\bf Third}, we find that the total number of attackers that are observed by the network telescope is dominated
by the number of attackers from a certain country $X$.\footnote{We were fortunate to see the real, rather than anonymized, attacker IP addresses, which allowed us to
aggregate the attackers based on their country code.
Our study was approved by IRB.} Moreover, we observe that both the number of attackers from country $X$ and the total number of attackers
exhibit a strong periodicity. Although we cannot precisely pin down the root cause of this {\em dominance and periodicity} phenomenon,
it does suggest that thoroughly examining the traffic between country $X$ and the rest of the Internet may significantly improve cybersecurity.
{\bf Fourth}, we investigate whether or not substantially smaller network telescopes would give approximately the same statistics that would be offered by a single, large network telescope.
This question is interesting on its own and, if answered affirmatively, could lead to more cost-effective operation of network telescopes.
Unfortunately, our analysis shows that substantially smaller telescopes might not be as useful a single, large telescope (of $2^{24}$ IP addresses).

\smallskip

\noindent{\bf Related Work.}
One approach to understanding cybersecurity posture is to analyze network telescope data.
Studies based on telescope data can be classified into two categories.
The first category analyzes telescope data {\em alone},
and the present study falls into this category.
These studies include the characterization of Internet background radiation \cite{Pang:2004:CIB,Wustrow:2010:IBR},
the characterization of scan activities \cite{Allman:2007:BHS},
and the characterization of backscatter for estimating global DOS activities \cite{Moore:2006:IID,Hussain:2003:FCD}.
However, we analyze cybersecurity posture, especially with regard to attacks that are likely caused by malicious worm, virus and bot activities.
This explains why we exclude the backscatter data (which is filtered as noise in the present paper).
The second category of studies analyzes telescope data together with other kinds of relevant data.
These studies include the use of telescope data and network-based intrusion detection and firewall logs to analyze Internet intrusion activities \cite{Yegneswaran:2003:IIG},
the use of out-of-band information to help analyze worm propagation \cite{moore:2002:code-red,moore:2003:slammer,bailey:2005:blaster}, and
the use of active interactions with remote IP addresses to filter misconfiguration-caused traffic \cite{Pang:2004:CIB}.
There are also studies that are somewhat related to ours, including the identification of one-way traffic
from data where two-way traffic is well understood \cite{Lee:2007:PMO,Allman:2007:BHS,Brownlee:2012:OTM,Treurniet:2011:NAC,Glatz:2012:CIO}.

The other approach to understanding cybersecurity posture is to analyze data collected by honeynet-like systems
(e.g., \cite{DBLP:conf/uss/Provos04,bailey:2005:blackhole,DBLP:journals/tifs/LiGCP11,DBLP:series/ais/BarfordCGLPY10,DBLP:journals/tifs/ZhanXX13}).
Unlike network telescopes, these systems can interact with remote computers and therefore allow for richer analysis,
including the automated generation of attack signatures \cite{Kreibich:2004:HCI:972374.972384,BarfordUsenixSecurity05}.

To the best of our knowledge, we are the first to formally define {\em cybersecurity posture} via three time series:
the number of victims, the number of attackers, and the number of attacks.

The rest of the paper is organized as follows.
Section \ref{sec:data-representation} describes the data and defines cybersecurity posture.
Section \ref{sec:statistical-preliminaries} briefly reviews some statistical preliminaries.
Section \ref{sec:victim-situation} defines and analyzes the sweep-time.
Section \ref{sec:attacker-situation} investigates the dominance and periodicity phenomenon exhibited by the number of attackers.
Section \ref{sec:inference} investigates whether substantially smaller network telescopes would be sufficient or not.
Section \ref{sec:limitations} discusses the limitations of the present study.
Section \ref{sec:conclusion} concludes the paper.

\section{Representation of Data and Definition of Cybersecurity Posture \label{sec:data-and-method}}
\label{sec:data-representation}

\noindent{\bf Data Description.}
The data we analyze was collected between 3/1/2013 and 3/31/2013 by CAIDA's network telescope,
which is a passive monitoring system based on a globally routeable but unused /8 network
(i.e., $1/256$ of the entire Internet IP v4 address space)  \cite{ucsd:caida}.
Since a network telescope passively collects unsolicited traffic, the collected traffic would contain {\em malicious traffic} that reaches the telescope
(e.g., automated malware spreading),
but may also contain {\em non-malicious traffic} --- such as
Internet background radiation (e.g., backscatter caused by the use of spoofed source IP addresses that happen to belong to the telescope)
and misconfiguration-caused traffic (e.g. mistyping an IP address by a remote computer).
This means that pre-processing the raw data is necessary.
At a high level, we will analyze data $D_1$ and $D_2$, which are sets of {\em flows}  \cite{ClaffyIEEEJSAC95} and
are obtained by applying the pre-processing procedures described below.

{\bf Data $D_1$.} Based on CAIDA's standard pre-processing \cite{ucsd:caida:flowtuple},
the collected IP packets are organized based on eight fields: source IP address, destination IP address,
source port number, destination port number, protocol, TTL (time-to-live), TCP flags, and IP length. The flows are reassembled from the IP packets and then classified
into three classes: {\em backscatter}, {\em ICMP request} and {\em other}. At a high level, backscatter traffic is identified via
TCP SYN+ACK, TCP RST, while ICMP request is identified via ICMP type 0/3/4/5/11/12/14/16/18.
(A similar classification method is used in \cite{Wustrow:2010:IBR}.)
Since (i) backscatter-based analysis of DOS attacks has been conducted elsewhere (e.g., \cite{Moore:2006:IID,Hussain:2003:FCD}), and (ii) ICMP has been mainly used
to launch DOS attacks (e.g., {\em ping flooding} and {\em smurf or fraggle} attacks
\cite{Moore:2006:IID,Weiler::2002:hddos,lau:2000:ddos}),
we disregard the traffic corresponding to {\em backscatter} and {\em ICMP request}.
Since we are more interested in analyzing cybersecurity posture corresponding to attacks that are launched through the TCP/UDP protocols,
we focus on the TCP/UDP traffic in the {\em other} category mentioned above.
We call the resulting data $D_1$, in which each TCP/UDP flow is treated as a distinct attack.

{\bf Data $D_2$.} Although (i) $D_1$ already excludes the traffic corresponding to {\em backscatter} and {\em ICMP request}, and (ii) $D_1$ only consists of TCP/UDP
flows in the {\em other} category mentioned above,
$D_1$ may still contain flows that are caused by misconfigurations.
Eliminating misconfiguration-caused flows in network telescope data is a hard problem because network telescope is passive (i.e., not interacting with
remote computers \cite{Gringoli:2009:GPU:1629607.1629610}).
Indeed, existing studies on recognizing misconfiguration-caused traffic had to use payload information (e.g., \cite{Zhichun:2011:P2P}),
which is however beyond the reach of network telescope data.
Note that recognizing misconfiguration-caused traffic is even harder than recognizing one-way traffic already
(because misconfiguration can cause both one-way {\em and} two-way traffic),
and that solving the latter problem already requires using extra information (such as
two-way traffic \cite{Lee:2007:PMO,Allman:2007:BHS,Brownlee:2012:OTM,Treurniet:2011:NAC,Glatz:2012:CIO}).
These observations suggest that we use some heuristics to filter probable misconfiguration-caused flows from $D_1$.
Our examination shows that, for example, 50\% (81\%) attackers launched 1 attack ($\leq 9$ attacks, correspondingly) against the telescope during the month.
We propose to extract $D_2$ by filtering from $D_1$
the flows that correspond to remote IP addresses that initiate fewer than 10 flows/attacks during the month.
This heuristic method filters possibly many, if not most, misconfiguration-caused flows in $D_1$.
Even though the ground truth (i.e., which TCP/UDP flows correspond to malicious attacks)
is not known, $D_2$ might be closer to the ground truth than $D_1$.

\smallskip

\noindent{\bf Data Representation.}
In order to analyze the TCP/UDP flow data $D_1$ and $D_2$,
we represent the flows through time series at some {\em time resolution} $r$.
We consider two time resolutions (because a higher resolution leads to more accurate statistics):
hour, denoted by ``$H$,'' and minute, denoted by ``$m$.''
For a given time resolution of interest, the total time interval $[0,T]$ is divided into short periods $[i,i+1)$ according to time resolution $r\in \{H, m\}$,
where $i=0,1,\ldots,T-1$, and $T=744$ hours (or $T=4,464$ minutes) in this case.
We organize the flows into time series from three perspectives:
\begin{itemize}
\item the number of {\em victims} (i.e., network telescope IP addresses that are ``hit'' by
remote attacking IP addresses contained in $D_1$ or $D_2$) per time unit at time resolution $r$,
\item the number of {\em attackers} (i.e., the remote attacking IP addresses contained in $D_1$ or $D_2$) per time unit at time resolution $r$, and
\item the number of {\em attacks} per time unit at time resolution $r$ (i.e., TCP/UDP flows initiated from remote attacking IP addresses in $D_1$ or $D_2$ are treated as attacks).
\end{itemize}

\begin{figure}[!hbtp]
\centering
\includegraphics[width=.7\textwidth]{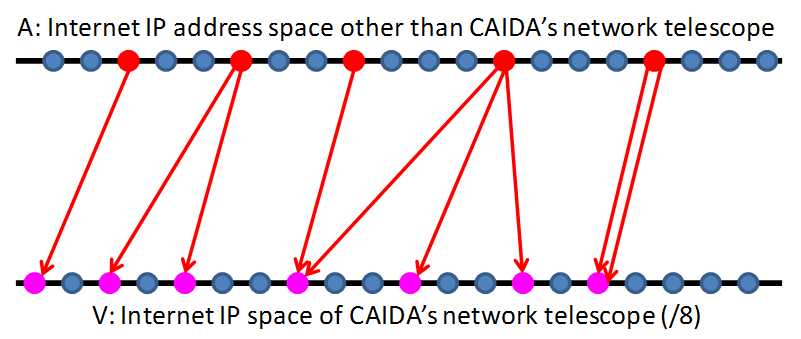}
\caption{Illustration of the attacker-victim relation during time interval $[i,i+1)$ at time resolution $r\in \{H,r\}$ in $D_1$:
each dot represents an IP address, a red-colored dot represents an attacking IP address (i.e., attacker),
a pink-colored dot represents a victim, each arrow represents an attack (i.e., TCP/UDP flow),
the number of attackers is $|A(r;i,i+1)|=5$, the number of victims is $|V(r;i,i+1)|=7$, and the number of attacks is $y(r;i,i+1)=9$.
The same holds for data $D_2$.
\label{fig:illustration}}
\end{figure}

As illustrated in Figure \ref{fig:illustration}, let $V$ be CAIDA's fixed set of telescope IP addresses,
and $A$ be the rest of IP addresses in cyberspace, where $|A|= 2^{32}-|V|$.
The major notations are (highlighted and) defined as follows:
\begin{itemize}
\item $V,A$: the set of CAIDA's network telescope IP addresses and the set of the rest IP v4 addresses, respectively.
\item $r\in\{H,m\}$: time resolution ($H$: per hour; $m$: per minute).
\item $V(r;i,i+1) \subseteq V$ and $V'(r;i,i+1) \subseteq V$: the sets of  {\em victims} attacked at least once during time interval $[i,i+1)$ at time resolution $r$ in
                          $D_1$ and $D_2$, respectively.
\item $V(r;i,j)=\bigcup_{\ell=i}^{j-1}V(r;\ell,\ell+1)$ and $V'(r;i,j)=\bigcup_{\ell=i}^{j-1}V'(r;\ell,\ell+1)$:
the cumulative set of victims that are attacked at some point during time interval $[i,j)$ at time resolution $r$ in data $D_1$ and $D_2$, respectively.
\item $V(r;0,T)$ and $V'(r;0,T)$: the sets of victims that are attacked at least once during time interval $[0,T)$ in $D_1$ and $D_2$, respectively. Note that these sets
        are actually independent of time resolution $r$, but we keep $r$ for notational consistence.
\item $A(r;i,i+1)\subseteq A$ and $A'(r;i,i+1)\subseteq A$: the sets of {\em attackers} that launched attacks against some $v\in V$
 during time interval $[i,i+1)$ at time resolution $r$ in $D_1$ and $D_2$, respectively.
\item $y(r;i,i+1)$ and $y'(r;i,i+1)$: the numbers of  {\em attacks} that are launched
against victims belonging to $V(r;i,i+1)$ and $V(r;i,i+1)$ during time interval $[i,i+1)$, respectively.
\end{itemize}

\noindent{\bf Cybersecurity Posture.}
We define cybersecurity posture as:
\begin{definition}
\emph{(cybersecurity posture)}
For a given time resolution $r$ and network telescope of IP address space $V$,
the {\em cybersecurity posture} as reflected by telescope data $D_1$
is described by the phenomena and (statistical) properties exhibited by the following three time series:
\begin{itemize}
\item the number of victims $|V(r;i,i+1)|$,
\item the number of attackers $|A(r;i,i+1)|$, and
\item the number of attacks $y(r;i,i+1)$,
\end{itemize}
where $i=0,1,\ldots$. Similarly, we can define cybersecurity posture corresponding to $D_2$.
\end{definition}
Based on the above definition of cybersecurity posture, the main research task is to characterize the phenomena and
statistical properties of the three time series (e.g., how can we predict them?) and the similarity between them.
As a first step, we characterize, by using data $D_1$ as an example,
the {\em number} of victims $|V(r;i,i+1)|$ rather than the {\em set} of victims $V(r;i,i+1)$,
and the {\em number} of attackers $|A(r;i,i+1)|$ rather than the {\em set} of attackers $A(r;i,i+1)$.
We leave the characterization of the {\em sets} of victims and attackers to future study.
Moreover, we characterize the {\em number} of attacks $y(r;i,i+1)$ rather than the specific classes of attacks,
because telescope data does not provide rich enough information to recognize specific attacks.
In Section \ref{sec:limitations}, we will discuss limitations of the present study, including the ones that are imposed by the heuristic data pre-processing method
for obtaining $D_1$ and $D_2$.

\section{Statistical Preliminaries}
\label{sec:statistical-preliminaries}

We briefly review some statistical concepts and models dealing with time series data,
while referring their formal descriptions and technical details to \cite{CC2008,engle1982,Tsay10,neter1996applied}.

\smallskip

\noindent{\bf Brief review of some statistical concepts.}
Time series can be described by statistical models, such as the AutoRegressive Integrated Moving Average (ARIMA) model and the
Generalized AutoRegressive Conditional Heteroskedasticity (GARCH) model that will be used in the paper.
The ARIMA model is perhaps the most popular class of time series models in the literature. It includes
many specific models,  such as random walk, seasonal trends, stationary, and non-stationary models \cite{CC2008}.
ARIMA models cannot accommodate high volatilities of time series data,
which however can be accommodated by GARCH models \cite{engle1982}.
GARCH models also can capture many phenomena,
such as  dynamic dependence in variance, skewness, and heavy-tails \cite{Tsay10}.

In order to find accurate models for describing time series data, we need to do model selection.
There are many model selection criteria, among which the Akaike's Information Criterion (AIC) is widely used.
This criterion is based on appropriately balancing between goodness of fit and model
complexity. It is defined in such a way that the smaller the AIC value, the better the model \cite{neter1996applied}.

\smallskip

\noindent{\bf Measuring the difference (or distance) between two time series}.
We need to measure the difference (or distance) between two time series: $Z_1,Z_2,\ldots$ and $Z'_1,Z'_2,\ldots$, where $Z_i\geq 0$ and $Z'_i\geq 0$ for $i=1,2,\ldots$.
This difference measure characterizes:
\begin{enumerate}
\item the {\em fitting} error, where the $Z_i$ time series may represent the observed values and the $Z'_i$ time series may represent the fitted values;
\item the {\em prediction} error, where the $Z_i$ time series may correspond to the observed values and the $Z'_i$ time series may correspond to the predicted values;
\item the {\em approximation} error, where the $Z_i$ time series may describe the observed values and the $Z'_i$ time series may describe the
values that may be inferred (i.e., estimated or approximated) from other data sources (e.g., we may want to know whether or not the statistics derived from
the data collected by a large network telescope can be inferred from the data collected by a much smaller network telescope).
\end{enumerate}
For conciseness, we use the  standard and popular measure known as Percent Mean Absolute Deviation (PMAD) \cite{armstrong2001}.
Specifically, suppose $Z_{t},Z_{t+1},\ldots,Z_{t+\ell}$ are given data,
and $Z'_t,Z'_{t+1},\ldots,Z'_{t+\ell}$ are the fitted (or predicted, or approximated) data.
The overall fitting (or prediction, or approximation) error (or the PMAD value) is defined as
$\frac{\sum_{j=t}^{t+\ell} |Z_j-Z'_j|}{\sum_{j=t}^{t+\ell} Z_j}.$
The closer to 0 the PMAD value, the better the fitting (or prediction, or approximation).
We note that our analysis is not bound to the PMAD measure, and it is straightforward to adapt our analysis to incorporate other measures of interest.

\smallskip

\noindent{\bf Measuring the shape similarity between two time series}.
Two time series may be very different from a measure such as the PMAD mentioned above,
but may be similar to each other in their shape (perhaps after some appropriate re-alignments).
Therefore, we may need to measure such shape similarity between two time series.
Dynamic Time Warping (DTW) is a method for this purpose.
Intuitively, DTW aims to align two time series that may have the same shape and, as a result, the similarity between two time series
can be captured by the notion of {\em warping path} (aka {\em warping function}).
The closer the warping path to the diagonal, the more similar the two time series.
We use the DTW algorithm in the $R$ software package, which implements the algorithm described in \cite{giorgino2009}.

\section{Characteristics of Sweep-Time}
\label{sec:victim-situation}

\begin{figure}[!hbtp]
\centering
\subfigure[$|V(H;i,i+1)|$ in $D_1$]{\includegraphics[width=.49\textwidth]{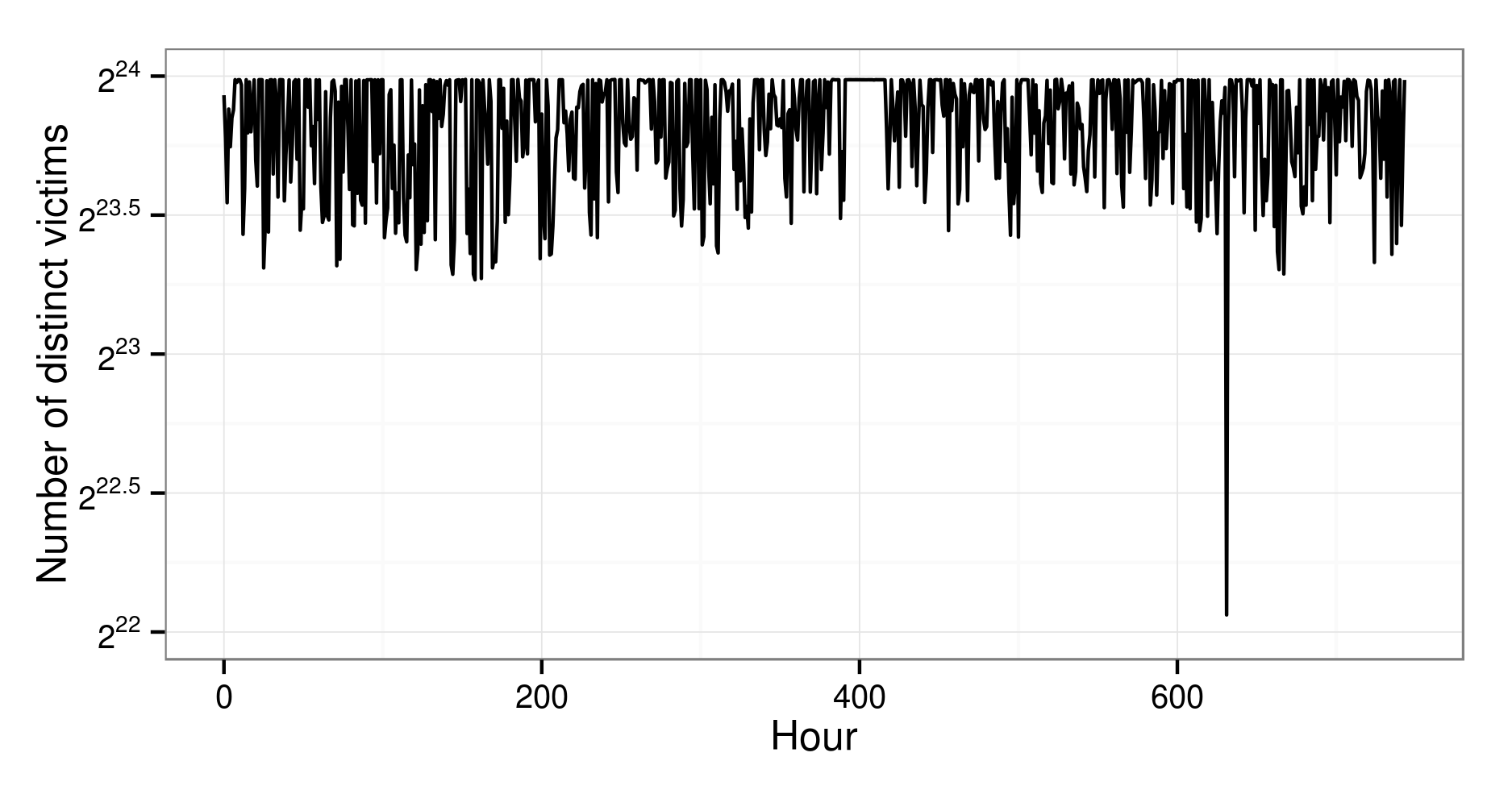} \label{fig:vic-uniq-vic-1}}
\subfigure[$|V'_1(H;i,i+1)|$ in $D_2$]{\includegraphics[width=.49\textwidth]{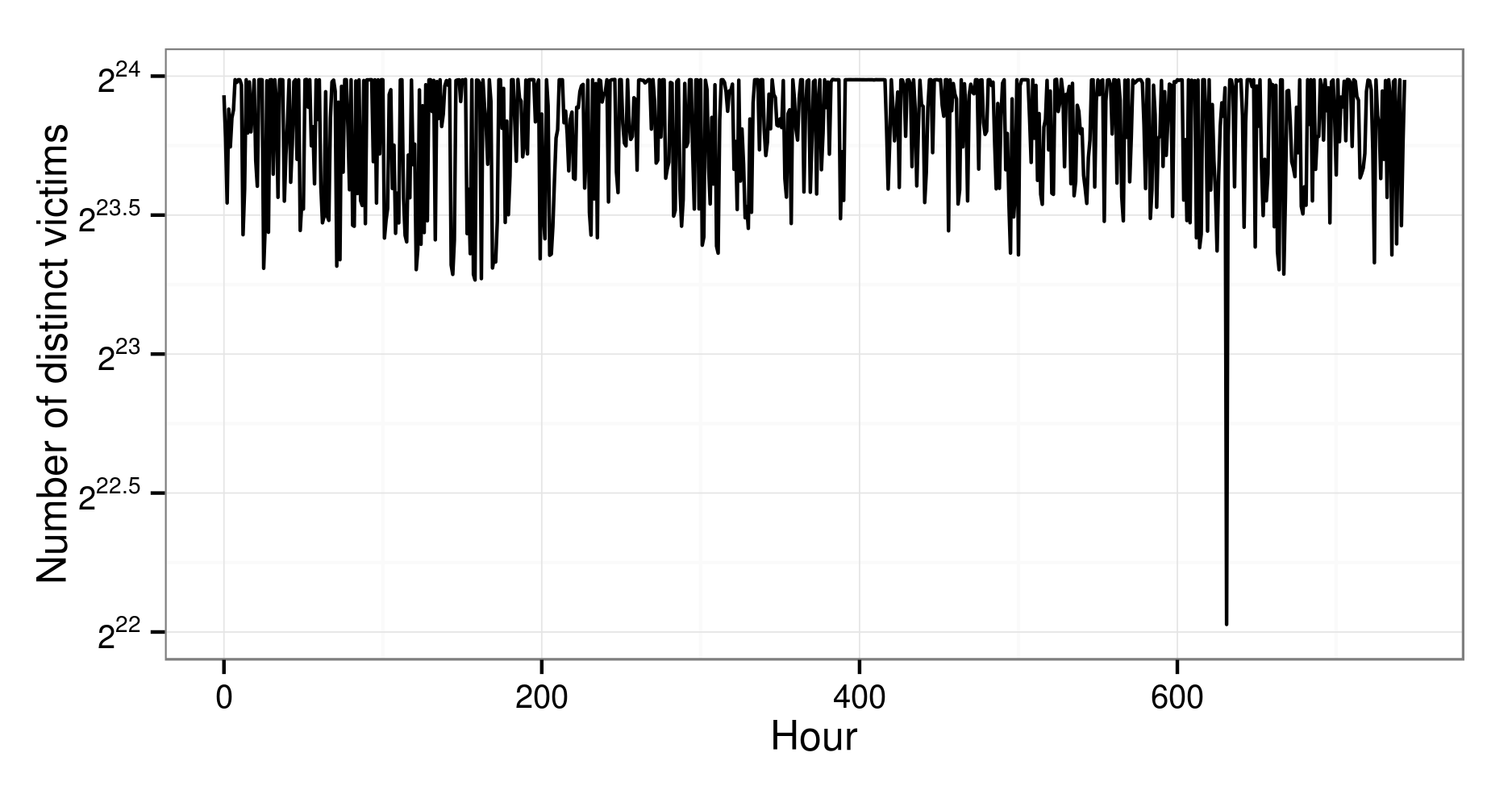} \label{fig:vic-uniq-vic-1-filtered}}
\caption{Time series of number of victims: $|V(H;i,i+1)|$ in $D_1$ and $|V'_1(H;i,i+1)|$ in $D_2$.
\label{fig:vic-uniq-vic-summary}}
\end{figure}

\noindent{\bf The Notion of Sweep-Time.}
Figure \ref{fig:vic-uniq-vic-summary}
describes the times series of $|V(H;i,i+1)|$ in $D_1$
and $|V'(H;i,i+1)|$ in $D_2$.
Using $D_1$ as example, we make the following observations (similar observations can be made for $D_2$).
First, there is a significant volatility at the 632nd hour, during which the number of victims is as low as $4,377,079\approx2^{22}$.
Careful examination shows that the total number of attackers
during the 632nd hour is very small, which would be the cause.
Second, most telescope IP addresses are attacked within a single hour.
For example, 15,998,907, or $96\%$ of $|V(H;1,733)|$, telescope IP addresses are attacked at least once during the first hour.
Third, no victims other than $V(H;0,703)$ are attacked during the time interval $[704,744)$.

Since Figure \ref{fig:vic-uniq-vic-summary} shows that there are a large number of victims per hour,
we ask the following question: How long does it take for most telescope IP addresses to be attacked at least once?
That is:
{\bf How long does it take for $\tau \times |V(H;0,T)|$ victims to be attacked at least once, where $0<\tau<1$?}
This suggests us to define the following notion of {\em sweep-time}, which is relative to the observation start time.
\begin{definition}
\emph{(sweep-time)}
With respect to $D_1$, the {\em sweep-time} starting at the $i$th time unit of time resolution $r$, denoted by $I_i$, is defined as:
$$\left|\bigcup_{\ell=i}^{I_i-1} V(r;\ell,\ell+1)\right|< \tau \times |V(H;0,T)| \leq \left|\bigcup_{\ell=i}^{I_i} V(r;\ell,\ell+1)\right|.$$
Corresponding to data $D_2$, we can define sweep-time $I'_i$ as:
$$\left|\bigcup_{\ell=i}^{I'_i-1} V'(r;\ell,\ell+1)\right|< \tau \times |V'(H;0,T)| \leq \left|\bigcup_{\ell=i}^{I'_i} V'(r;\ell,\ell+1)\right|.$$
\end{definition}
By taking into consideration the observation starting time $i$, we naturally obtain two time series of sweep-time:
 $I_0,I_1,\ldots$ for $D_1$ and $I'_0,I'_1,\ldots$ for $D_2$.
We want to characterize these two time series of sweep-time.

\smallskip

\begin{figure}[!hbtp]
\centering
\subfigure[Sweep-time in $D_1$\label{fig:vic-kill-1m-ts}]{\includegraphics[width=.49\textwidth]{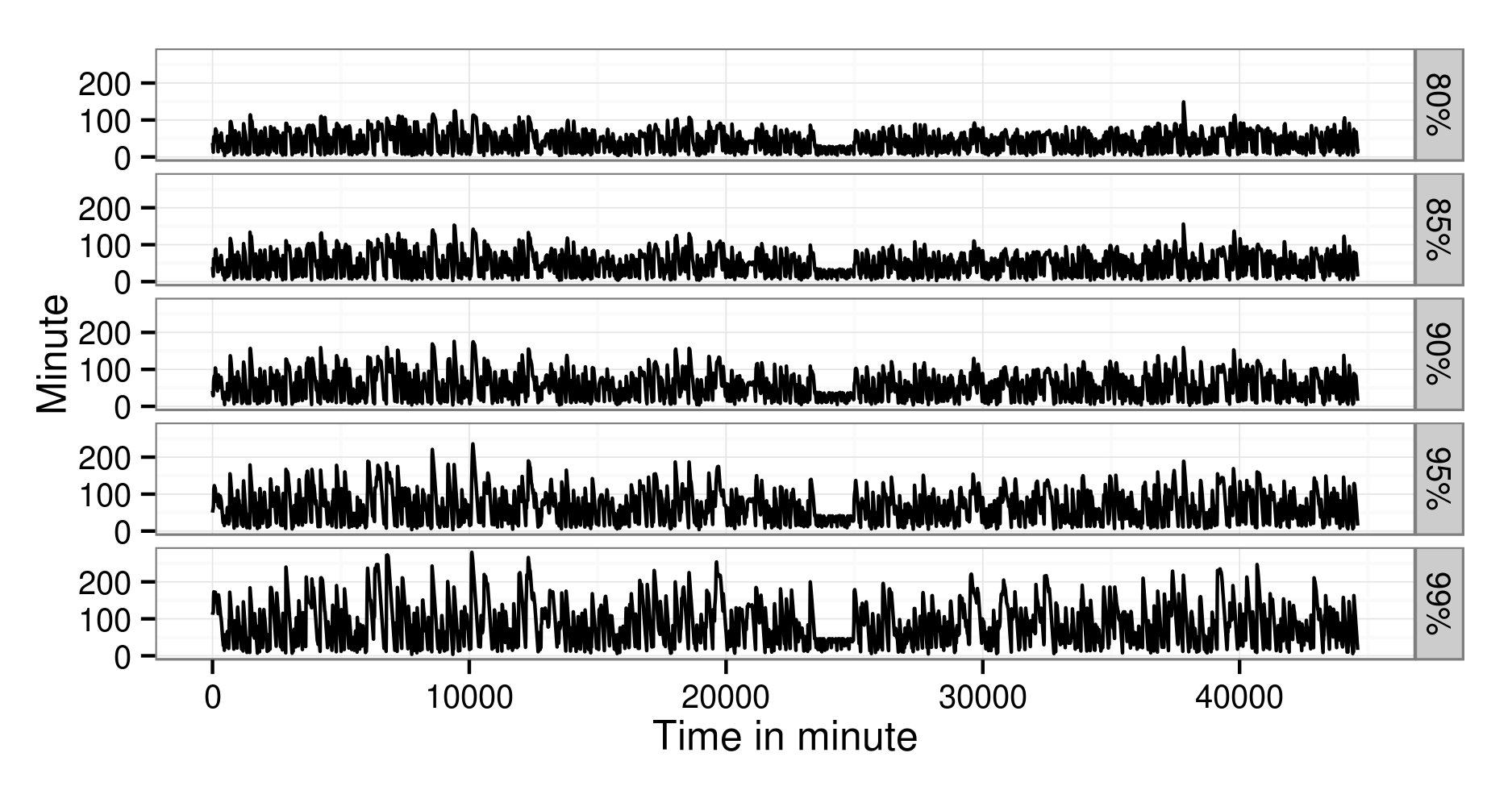}}
\subfigure[Sweep-time in $D_2$\label{fig:vic-kill-1m-ts}]{\includegraphics[width=.49\textwidth]{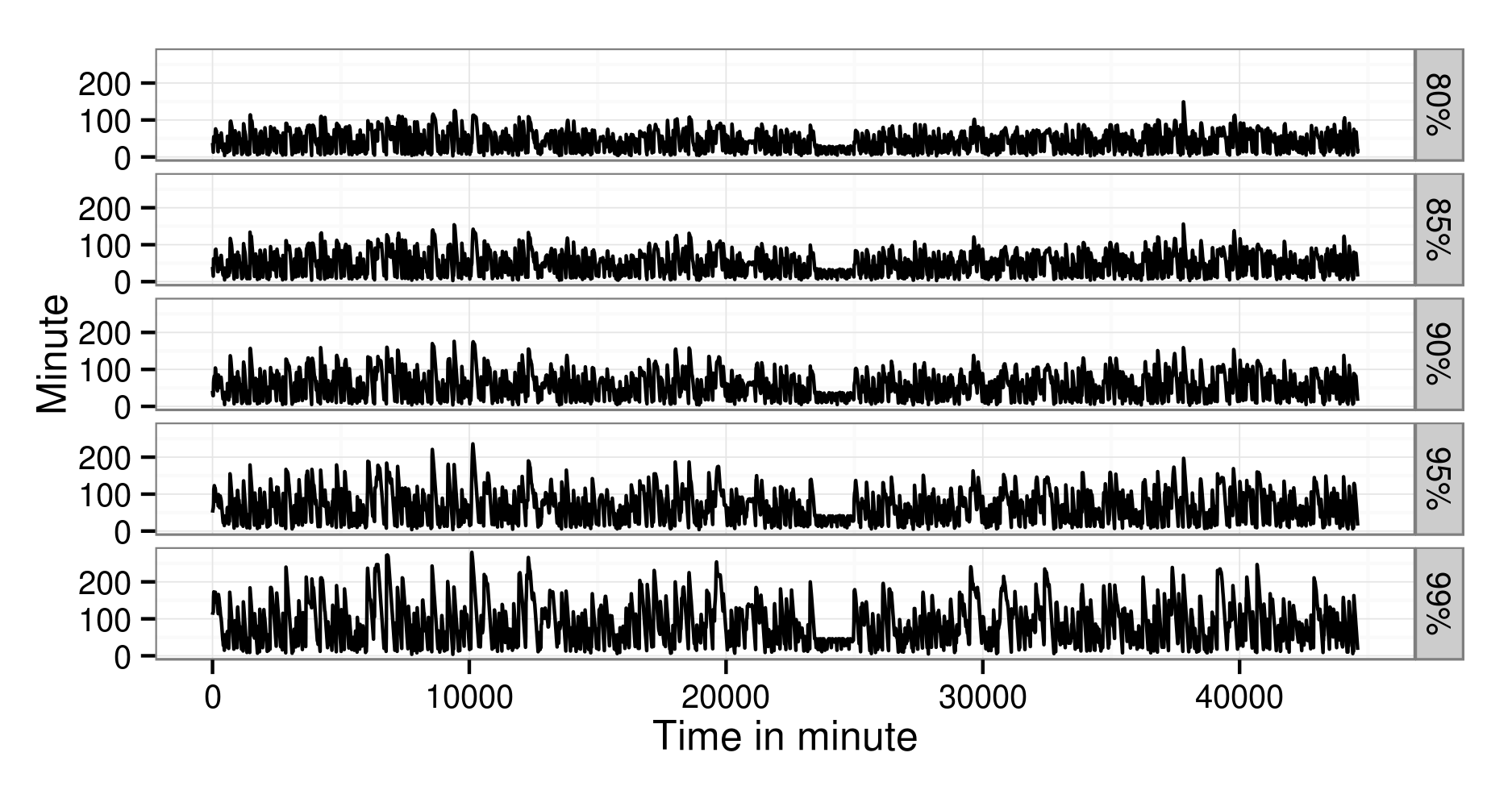}}
\caption{Time series plots of sweep-time ($y$-axis) with respect to $\tau \in\{80\%, 85\%, 90\%, 95\%,  99\%\}$,
where the $x$-axis represents the observation starting time that is sampled at every 10 minutes.
In other words, the plotted points are the sample $(0,I_0),(10,I_{10}),(20,I_{20}),\ldots$ rather than $(0,I_0),(1,I_1),\ldots$.
\label{fig:vic-killtime}}
\end{figure}

\noindent{\bf Characterizing Sweep-Time.}
Since  {\em per-hour} time resolution gives a coarse estimation of sweep-time,
we use  {\em per-minute} time resolution for better estimation of it.
Figure \ref{fig:vic-killtime} plots the time series of sweep-time $I_0,I_{10},I_{20},\ldots$,
namely a sample of $I_0,I_1,\ldots,I_{10},\ldots,I_{20},\ldots$ because it is too time-consuming to consider the latter entirely (time resolution: minute).

Figure \ref{fig:vic-killtime} suggests that the sweep-time time series exhibit similar shape.
Accordingly, we use the DTW method to characterize their similarity.
Recall that DTW aims to align two time series via
the notion of {\em warping path},
such that the closer the warping path to the diagonal, the more similar the two time series.

\begin{figure*}[hbtp]
\centering
\subfigure[{\tiny $D_1$ vs. $D_2$: $\tau=.80$}]{\includegraphics[width=.31\textwidth]{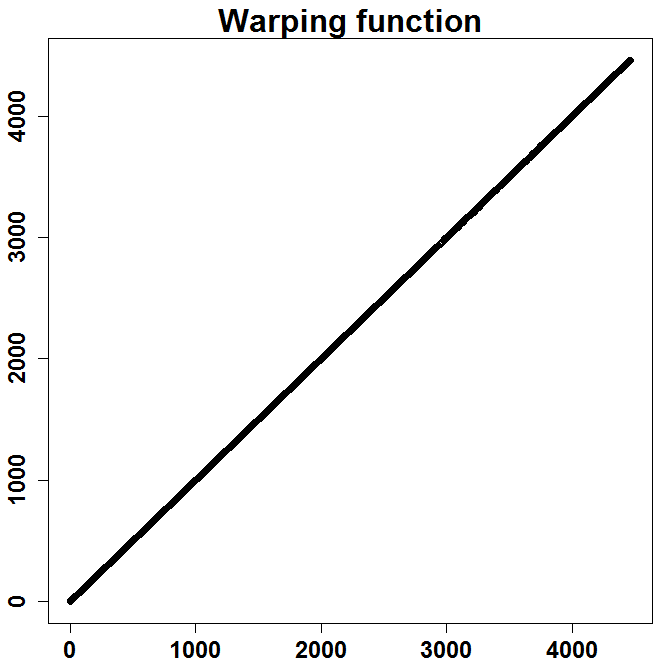}\label{fig:dtw-d1-d2-time-08}}
\subfigure[{\tiny $D_1$ vs. $D_2$: $\tau=.99$}]{\includegraphics[width=.31\textwidth]{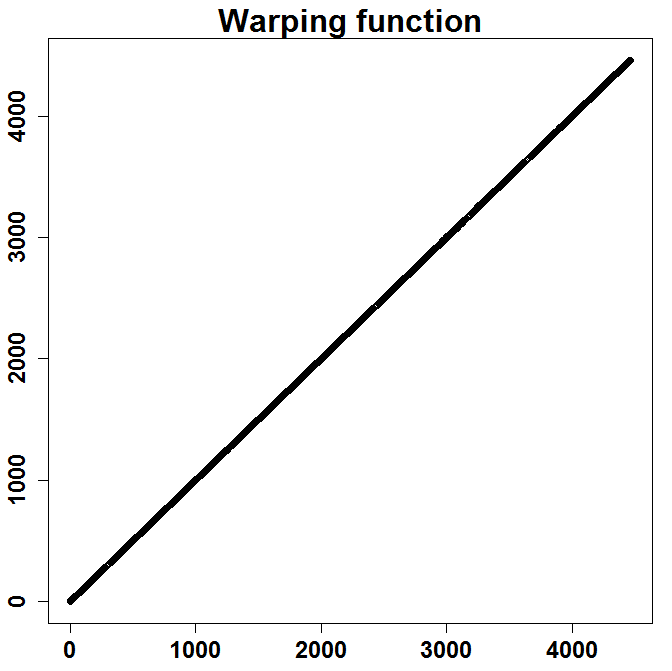}\label{fig:dtw-d1-d2-time-99}}
\subfigure[{\tiny $\tau=.99$ vs. $\tau=.80$: $D_1$}]{\includegraphics[width=.31\textwidth]{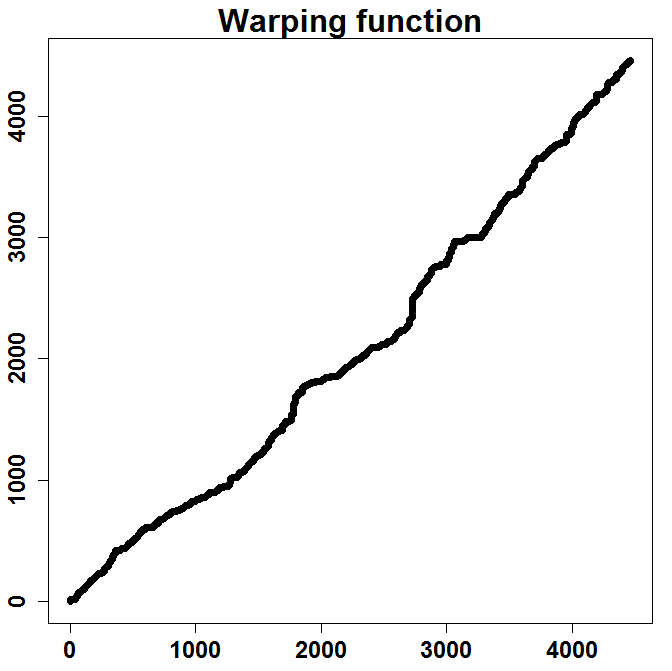}\label{fig-d1-time-99-80}}
\subfigure[{\tiny $\tau=.99$ vs. $\tau=.95$: $D_1$}]{\includegraphics[width=.31\textwidth]{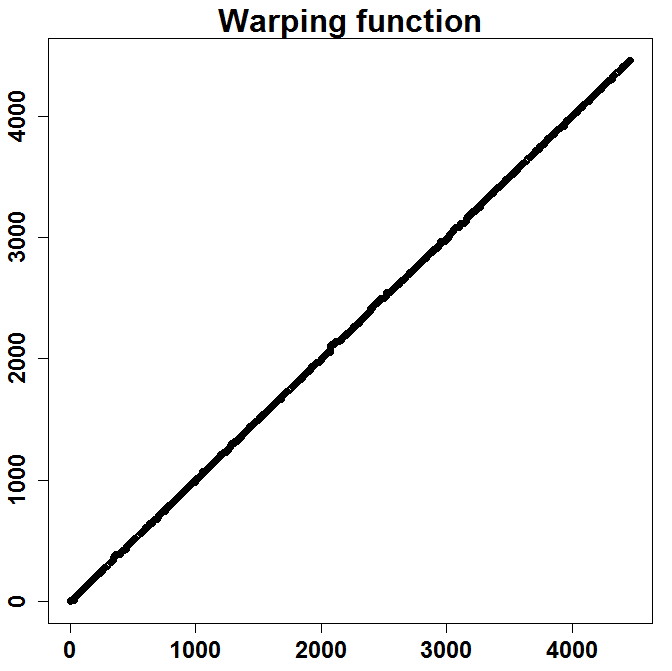}\label{fig:d1-time-99-95}}
\subfigure[{\tiny $\tau=.99$ vs. $\tau=.80$: $D_2$}]{\includegraphics[width=.31\textwidth]{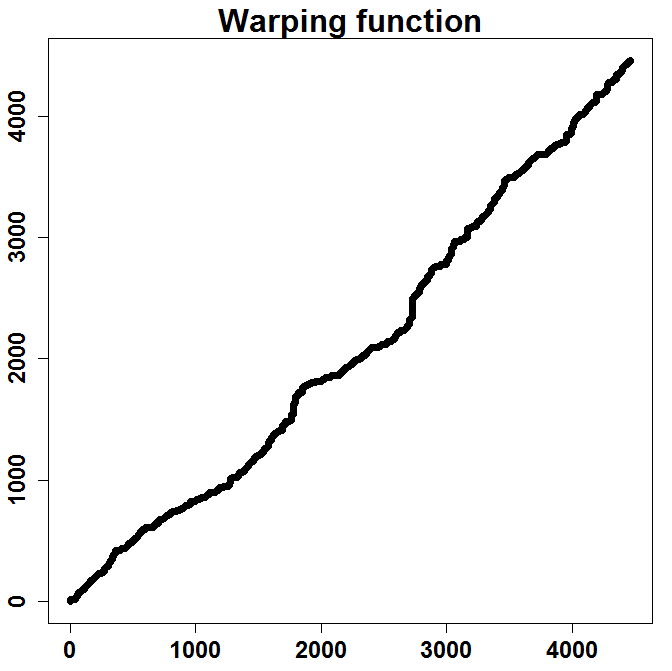}\label{fig:d2-time-99-80}}
\subfigure[{\tiny $\tau=.99$ vs. $\tau=.95$: $D_2$}]{\includegraphics[width=.31\textwidth]{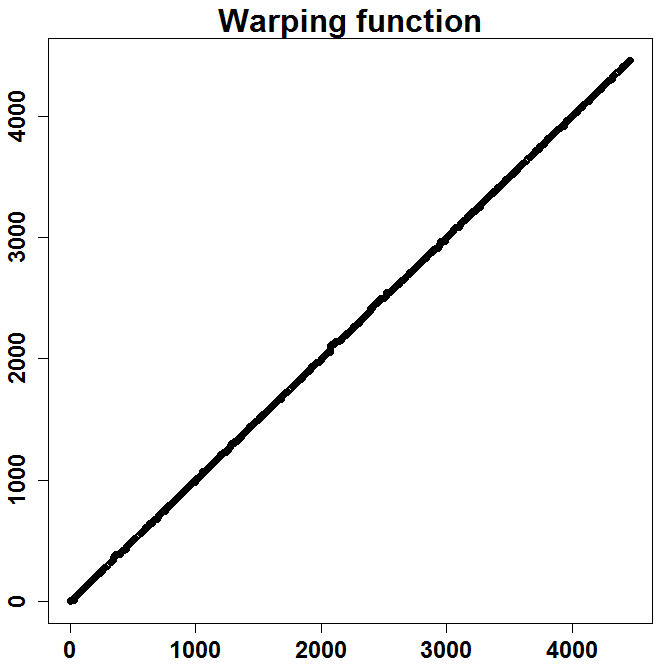}\label{fig:d2-time-99-95}}
\caption{DTW-based similarities of sweep-time time series with different threshold $\tau$.
\label{fig:sweep-time-dtw-d1-d2}}
\end{figure*}

Figure \ref{fig:sweep-time-dtw-d1-d2} confirms the above observation, by presenting some examples of the warping paths (the others are omitted due to space limitation).
Specifically, Figures \ref{fig:dtw-d1-d2-time-08} and \ref{fig:dtw-d1-d2-time-99} show that the sweep-times with respect to $\tau=.80$ and $\tau=.99$
are almost the same in $D_1$ and $D_2$, respectively.
Figures \ref{fig-d1-time-99-80} and \ref{fig:d1-time-99-95} show that for two thresholds, say $\tau_1$ and $\tau_2$,
the smaller the $|\tau_1-\tau_2|$, the more similar the two respective time series of sweep-time in $D_1$.
Figures \ref{fig:d2-time-99-80} and \ref{fig:d2-time-99-95} demonstrate the same phenomenon for $D_2$.

The above discussion suggests that the notion of sweep-time is not sensitive to spatial threshold $\tau$.
This leads us to ask: {\em\bf What is the distribution of sweep-time?}
However, this question makes sense only when the time series is stationary.
By using an augmented Dickey-Fuller test \cite{Tsay10}, we conclude that the sample of the sweep-time time series,
namely $I_0,I_{10},I_{20},\ldots$, is {\em not} stationary, which means
that time series $I_0,I_1,I_2\ldots$ is not stationary (otherwise, the sample should be stationary).
Therefore, we cannot use a single distribution to characterize the sweep-time;
Instead, we have to characterize the sweep-time as a stochastic process.
In order to identify good time series models that can fit the sweep-time,
we need to identify some statistical properties that are exhibited by sweep-time.
In particular, we need to know if the sweep-time is heavy-tailed, meaning that
the sweep-times greater or equal to $x_{\min}$ exhibit the power-law distribution,
where $x_{\min}$ is called {\em cut-off} parameter.

\begin{table}[htbp!]
\centering
\begin{tabular}{|c||r|r|r|r|r||c||r|r|r|r|r|}
  \hline
   $\tau$& $\alpha$ & $x_{\min}$ &  $KS$ & $p$-value & $\#\geq x_{\min}$ &    $\tau$& $\alpha$ & $x_{\min}$ &  $KS$ & $p$-value & $\#\geq x_{\min}$\\\hline
   \multicolumn{6}{|c|}{Dataset $D_{1}$ with time resolution 1-minute}   &   \multicolumn{6}{|c|}{Dataset $D_{2}$ with time-resolution 1-minute}\\\hline
   80\%  & 7.89  &  78  &  .05  & .14 &   475                            & 		80\%     & 8.46  & 82   &  .05   &  .19  &   391\\\hline
   85\%  & 8.46  &  94  &  .04  & .52 &   385 				 & 		85\%     & 8.37  & 95   &  .04   &  .36  &   379\\\hline
   90\%  & 8.89  &  118 &  .06  & .42 &   244                            & 		90\%     & 9.24  & 120  &  .05   &  .39  &   237\\\hline
   95\%  & 9.52  &  148 &  .05  & .68 &   193                            &		95\%     & 12.82 & 170  &  .04   &  .99  &   72\\\hline
   99\%  & 13.67 &  215 &  .04  & .98 &   131                            &		99\%     & 15.23 & 224  &  .04   &  .99  &   94\\\hline
\end{tabular}
\caption[Power-law test statistics]{Power-law test statistics of the sweep-time with respect to {\em spatial threshold} $\tau\in\{80\%, 85\%, 90\%, 95\%, 99\%\}$,
where $\alpha$ is the fitted power-law exponent,
$x_{\min}$ is the cut-off parameter,
$KS\in [0.04,0.06]$ is the Kolmogorov-Smirnov statistic \cite{clauset2009power} for comparing the fitted power-law distribution and the data (meaning that the fitting is good)
as indicated by that the $p$-values are $>>0.05$, and ``$\#\geq x_{\min}$''
represents the number of sweep-times that are greater than or equal to $x_{\min}$ (i.e., the number of sweep-times that are used for fitting).
\label{tbl:killtime-power-law}}
\end{table}

Table \ref{tbl:killtime-power-law} summarizes the power-law test statistics of the sweep-time with cut-off parameter $x_{\min}$.
We observe that for both $D_{1}$ and $D_{2}$, all the $\alpha$ values (i.e., the fitted power-law exponents) are very large.
For spatial threshold $\tau=80\%$ in $D_{1}$, we have $x_{\min}=78$ minutes,
meaning that the number of sweep-times that are greater than or equal to $x_{\min}$ is 475 (or 10.6\% out of 4,462).
As spatial threshold $\tau$ increases, $x_{\min}$ increases and the number of sweep-times greater than or equal to $x_{\min}$ decreases.
We also observe that for the same $\tau$, $D_{1}$ and $D_{2}$ have similar $x_{\min}$ values,
which means that the filtered attack traffic in $D_{1}$ does not affect the power-law property of the data.

\begin{figure}[!hbtp]
\centering
\subfigure[Fitting sweep-time in $D_1$.\label{fig:sweep-1m-ts-d1}]{\includegraphics[width=.46\textwidth]{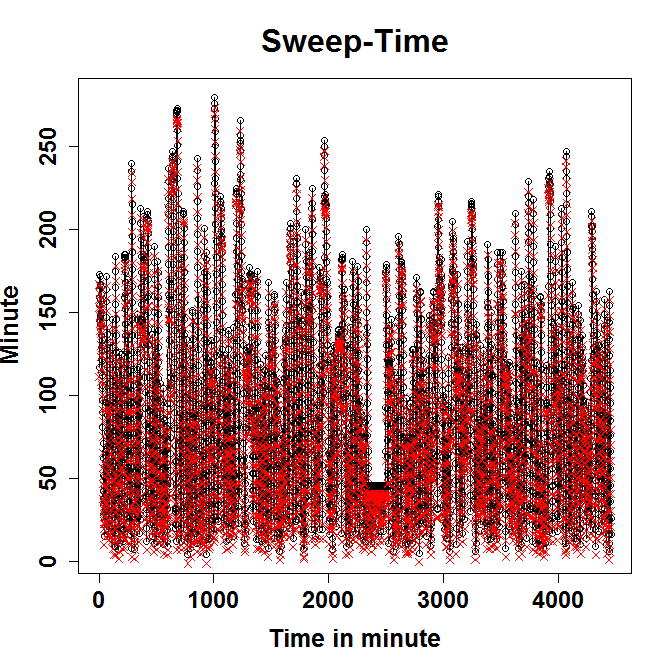}}
\subfigure[Fitting sweep-time in $D_2$.\label{fig:sweep-1m-ts-d2}]{\includegraphics[width=.46\textwidth]{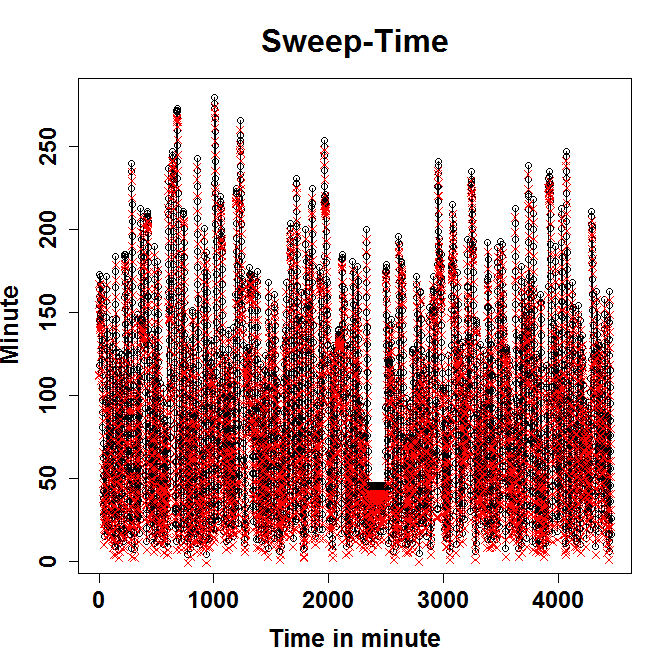}}
\caption{Fitting sweep-time with spatial threshold $\tau=.99$ (time resolution: minute), where black-colored dots are observed sweep-time values,
red-colored dots are fitted values.
\label{fig:fitting-sweep-time}}
\end{figure}

The above analysis suggests that in order to fit the sweep-time, we should use a model that can accommodate the power-law property.
Therefore, we use the ARMA+GARCH model, where ARMA accommodates the stable sweep-times smaller than $x_{\min}$,
and GARCH, with skewed student $t$-distribution, accommodates the power-law distributed sweep-times (which are greater than or equal to $x_{\min}$).
Consider spatial threshold $\tau=.99$ as an example.
Figures \ref{fig:sweep-1m-ts-d1} and \ref{fig:sweep-1m-ts-d2} plot the observed data and the fitting model for sweep-time $I_t$ (observation starting time $t$): 
$$
I_t-\mu_t=\phi_1(I_{t-1}-\mu_{t-1})+\phi_2(I_{t-2}-\mu_{t-2})+\epsilon_t,
$$
where $\mu_t=\mu+\xi\sigma_t$ is the dynamic mean composed of a constant term $\mu$ and standard deviation $\sigma_t$ of the error term,
$\sigma_t^2=\omega+\alpha_1 \epsilon_{t-1}+\beta_1 \sigma^2_{t-1}$,
$\mu_t =E[I_t]$, $\epsilon_t$ is the error term at time $t$, and
$\sigma_t^2=E(y_t-\mu_t)^2$ is the variance
modeled via the standard GARCH(1,1) process.
The fitting errors (PMAD values) are .121 and .119 for $D_1$ and $D_2$, respectively.

It was known that malware can infect almost all susceptible computers within a very short period of time (e.g., the Slammer worm \cite{moore:2003:slammer}),
meaning that the sweep-time with respect to {\em a specific observation starting time} is very small.
This is, in a sense, re-affirmed by our study.
However, for continuous attacks that are based on a bag of attacking tools, sweep-time should be better modeled with respect to {\em any} (rather than
a specific) observation starting time. We are the first to show that the sweep-time {\em cannot} be modeled by a random variable
(which would make the model in question easier to analyze though).
This leads to the following insight, which could guide future development of advanced cybersecurity models.
\begin{insight}
When one needs to model the sweep-time (i.e., the time it takes for each IP address of a $\tau$-portion of a large network space to be
attacked at least once), it should be modeled by a stochastic process rather than a random variable.
\end{insight}

\section{A Phenomenon Exhibited by Attacking IP Addresses}
\label{sec:attacker-situation}

\begin{figure}[hbtp!]
\centering
\subfigure[Origins of attackers in $D_1$]{\includegraphics[width=.36\textwidth]{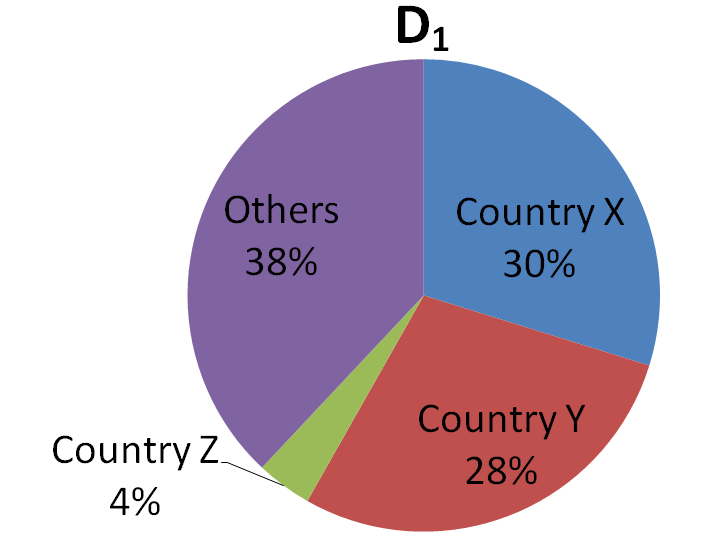} \label{fig:D1-top3-atkerno-pie}}
\subfigure[Origins or attackers in $D_2$]{\includegraphics[width=.36\textwidth]{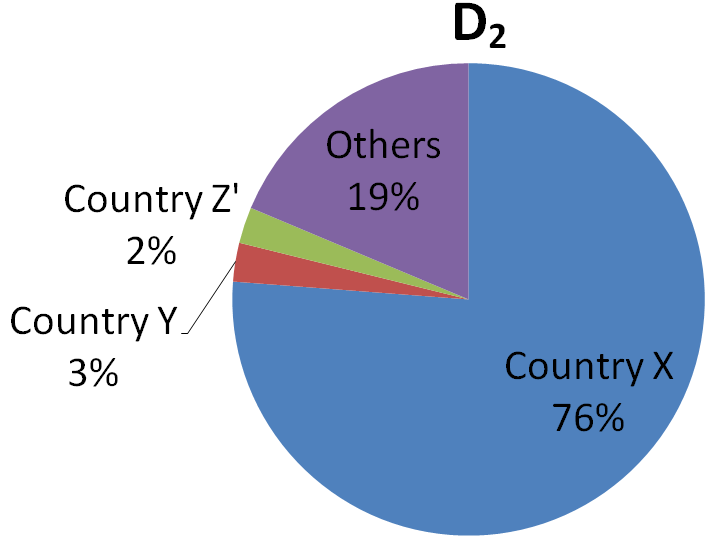} \label{fig:D2-top3-atkerno-pie}}
\caption{During the month, three countries, which are anonymized as $X$, $Y$ and $Z$, contribute to most of the attackers in $D_1$;
whereas countries $X$, $Y$ and $Z'$ ($Z'\neq Z$) contribute to most of the attackers in $D_2$.
\label{fig:pie-graph-attackers}}
\end{figure}

For each attacker IP address, we can use the WHOIS service to retrieve its country code.
Figures \ref{fig:pie-graph-attackers} plots the origins of attackers that contribute to most of the attackers (per country code).
Note that the category ``others'' in $D_1$ include 6,894,900 attacker IP addresses (or $1.7\%$ of the total number of attackers) whose country codes cannot be retrieved from the WHOIS service.
The category ``others'' in $D_2$ include 10,740 attacker IP addresses (or $0.01\%$ of the total number of attackers) whose country codes cannot be retrieved from the WHOIS service.
This means that many attacker IP addresses whose country codes cannot be retrieved from the WHOIS service are filtered.
Moreover, the attacker IP addresses with no country code do not have a significant impact on the result.
We observe that country $X$ contributes $30\%$ of the attackers in $D_1$ and $76\%$ of the attackers in $D_2$.
Country $Y$ contributes $28\%$ of the attackers in $D_1$ and $3\%$ of the attackers in $D_2$.
This is caused by the fact that $50\%$ attackers from country $X$ and $98\%$ attackers from country $Y$ launch fewer than 10 attacks during the month,
and therefore do not appear in $D_2$.
This prompts us to study the relationship between two time series: the total number of attackers and the number of attackers from country $X$.

\begin{figure}[!hbtp]
\centering
\subfigure[{\scriptsize Total \# of attackers vs. \# of attackers from $X$: $D_1$}\label{fig:vic-uniq-atker-cc-2}]{\includegraphics[width=.49\textwidth]{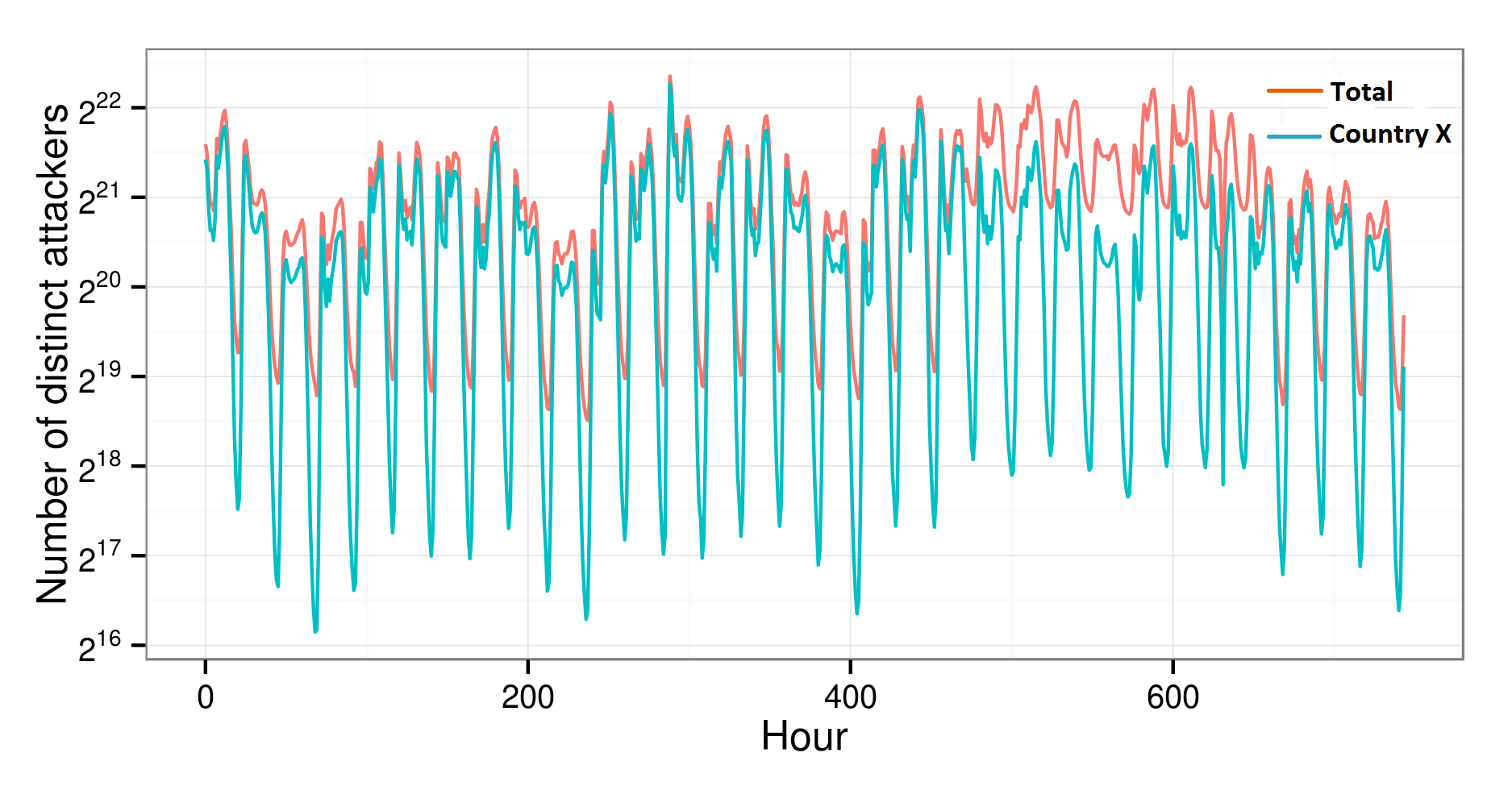}}
\subfigure[{\scriptsize Total \# of attackers vs. \# of attackers from $X$: $D_2$}\label{fig:D2-vic-uniq-atker-cc-2}]{\includegraphics[width=.49\textwidth]{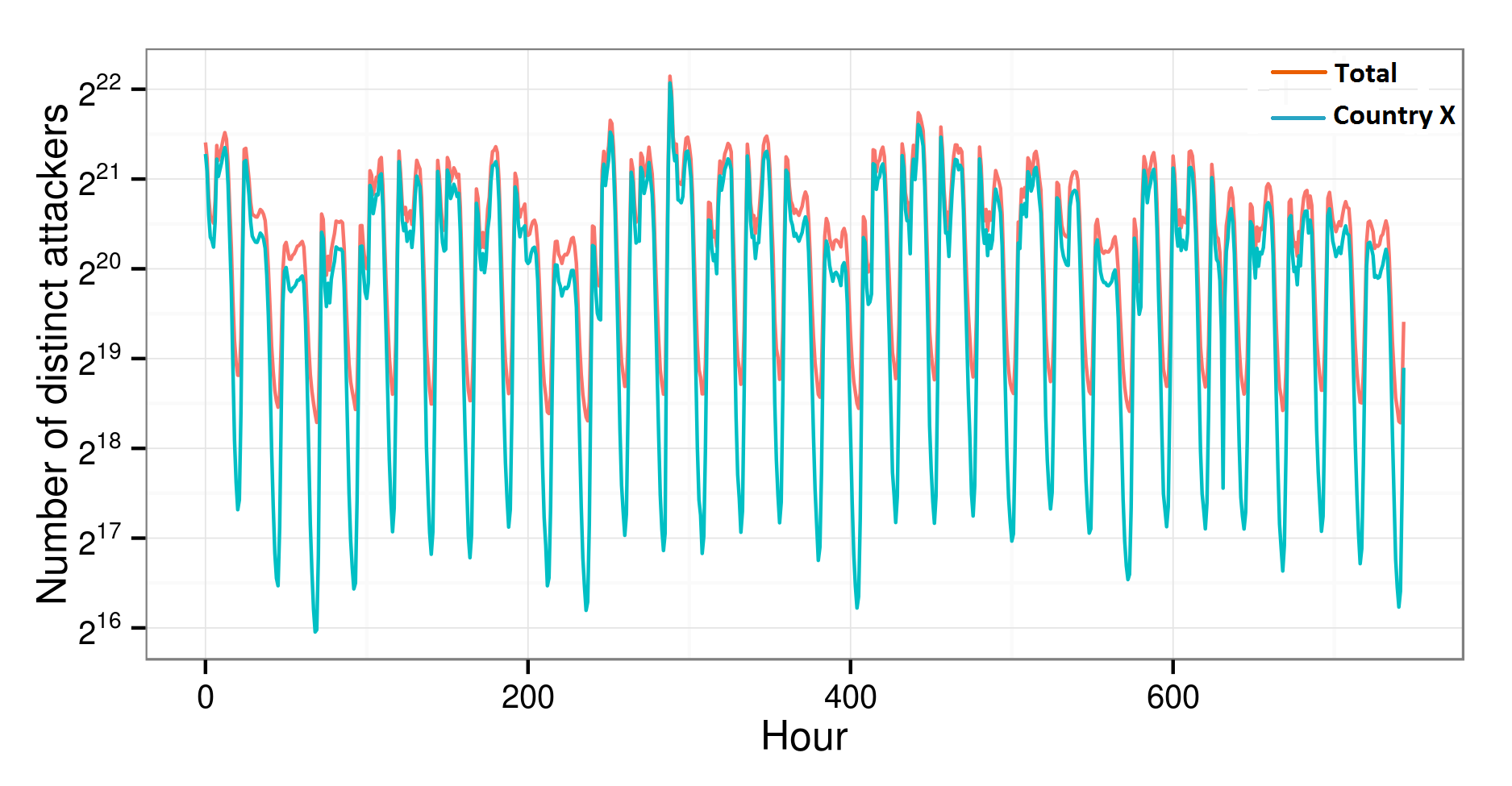}}
\caption{The dominance and periodicity phenomenon exhibited by two time series: the total number of attackers
versus the number of attackers from country $X$ (time resolution: hour).
\label{fig:vic-uniq-atker-stat-general}}
\end{figure}

\noindent{\bf The Dominance and Periodicity Phenomenon.}
Figure \ref{fig:vic-uniq-atker-stat-general}
compares the times series of the total number of attackers observed by the telescope
and the time series of the number of attackers from country $X$ in $D_1$ and $D_2$, respectively.
For $D_1$, Figure \ref{fig:vic-uniq-atker-cc-2} shows that the total number of attackers
during time interval $[455,630]$, namely the 176 hours between the 455th hour (on March 19, 2013) and the 630th hour (on March 27, 2013),
is substantially greater than its counterpart during the other hours.
This is caused by the substantial increase in the number of attackers from country $Y$,
despite that we do not know the root cause behind the substantial increase of attackers in country $Y$.
For $D_2$, Figure \ref{fig:D2-vic-uniq-atker-cc-2} does not exhibit the same kind of substantial increase during the interval $[455,630]$,
meaning that many of the ``emerging'' attackers from country $Y$ are filtered
(because they launched fewer than 10 attacks during the month).

\begin{figure}[!hbtp]
\centering
\subfigure[{\scriptsize Total number of attackers in $D_1$}\label{fig:ts-uniqAtker-waveBase}]{\includegraphics[width=.49\textwidth]{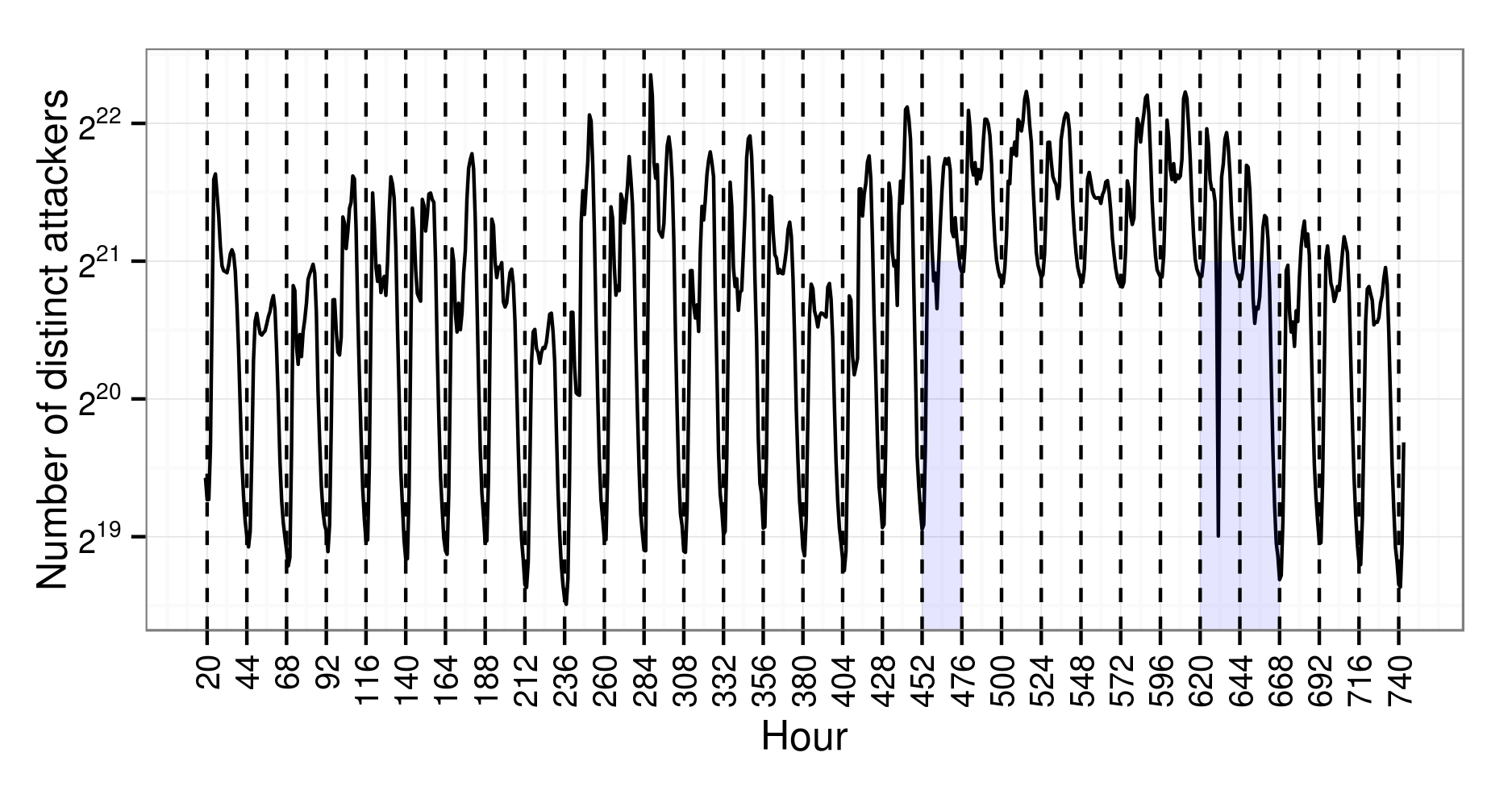}}
\subfigure[{\scriptsize Number of attackers from country $X$ in $D_1$}\label{fig:ts-uniqAtker-waveBase-CN}]{\includegraphics[width=.49\textwidth]{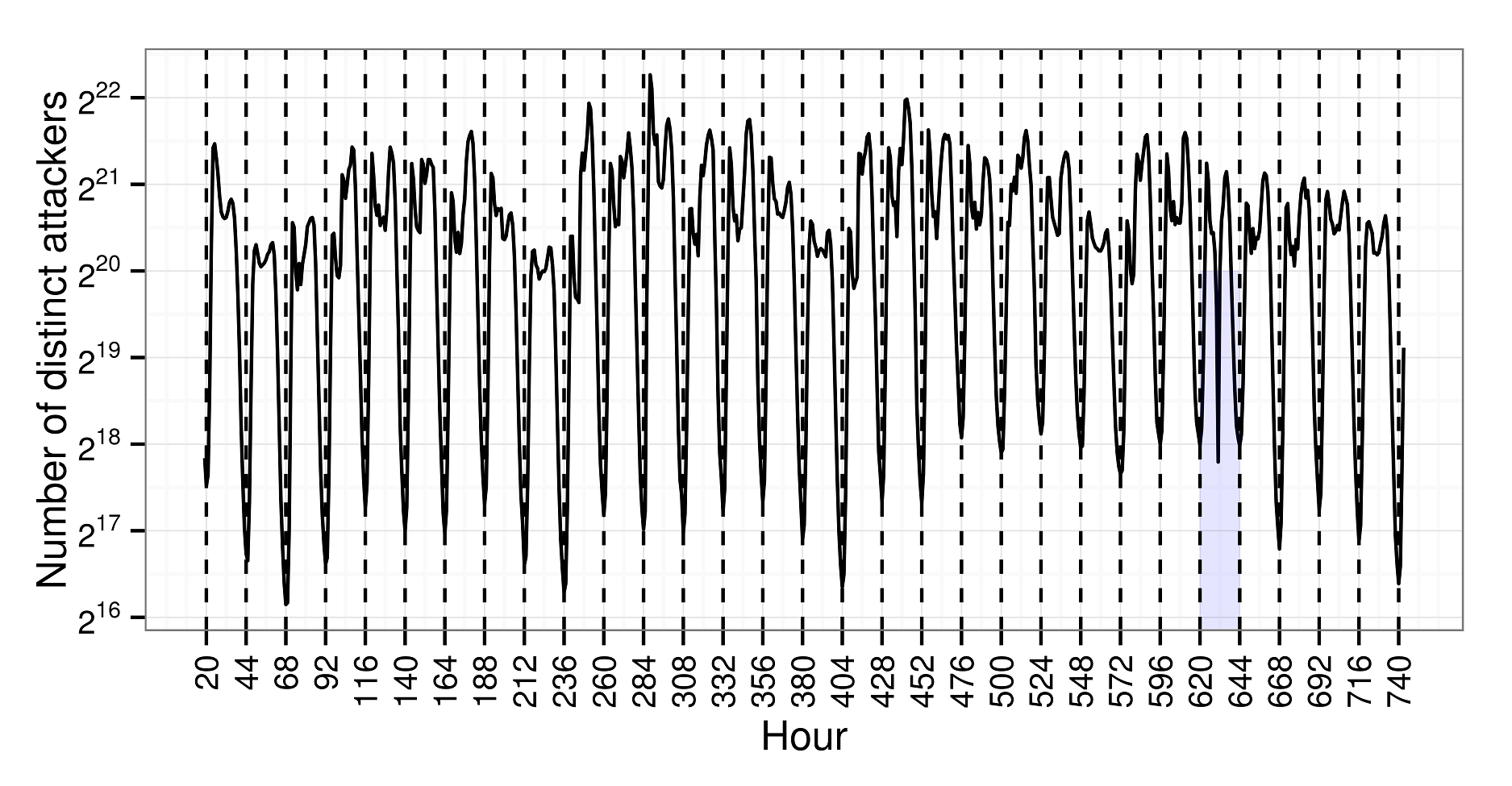}}
\subfigure[{\scriptsize Total number of attackers in $D_2$}\label{fig:D2-ts-uniqAtker-waveBase}]{\includegraphics[width=.49\textwidth]{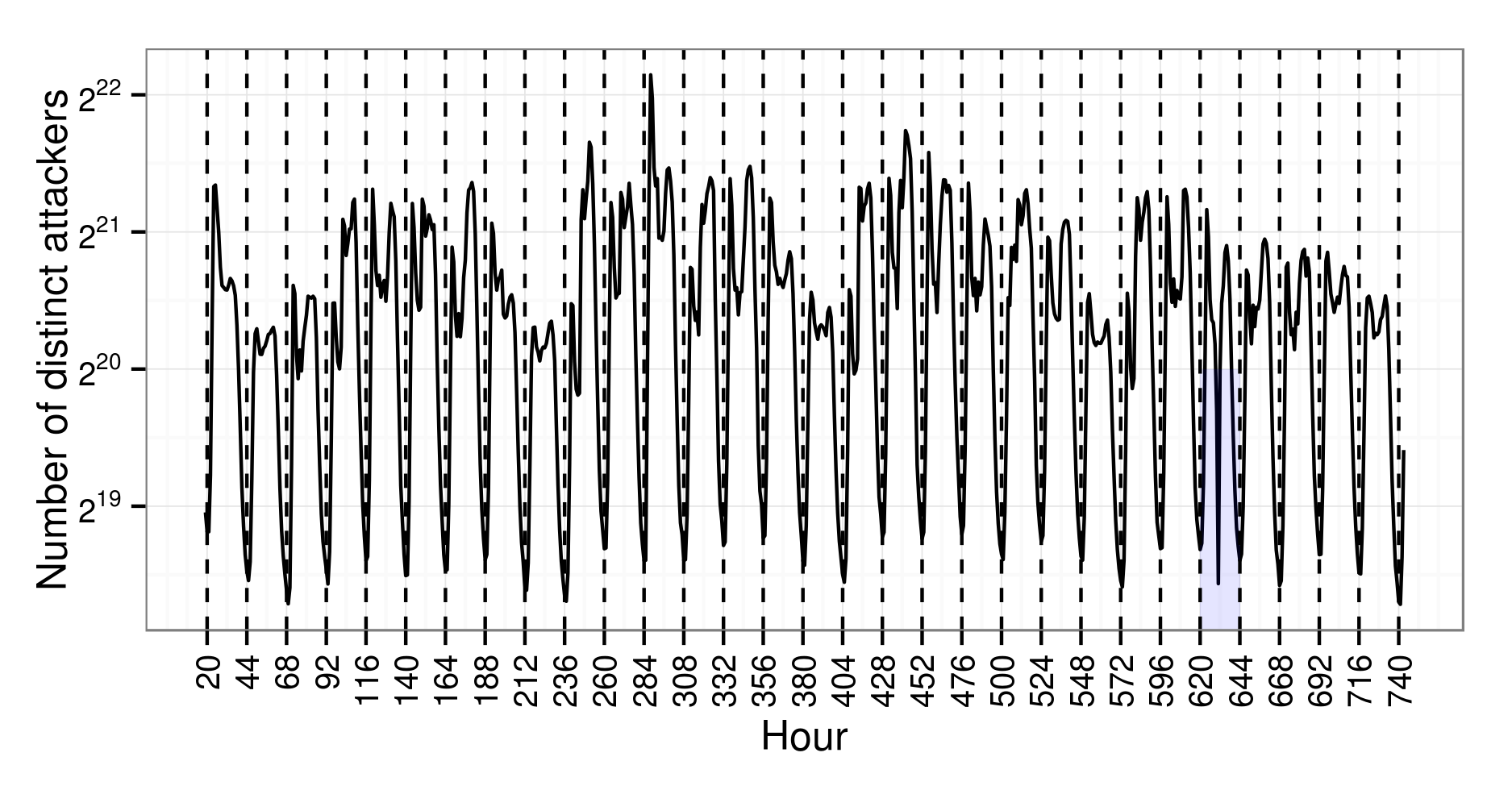}}
\subfigure[{\scriptsize Number of attackers from country $X$ in $D_2$}\label{fig:D2-ts-uniqAtker-waveBase-CN}]{\includegraphics[width=.49\textwidth]{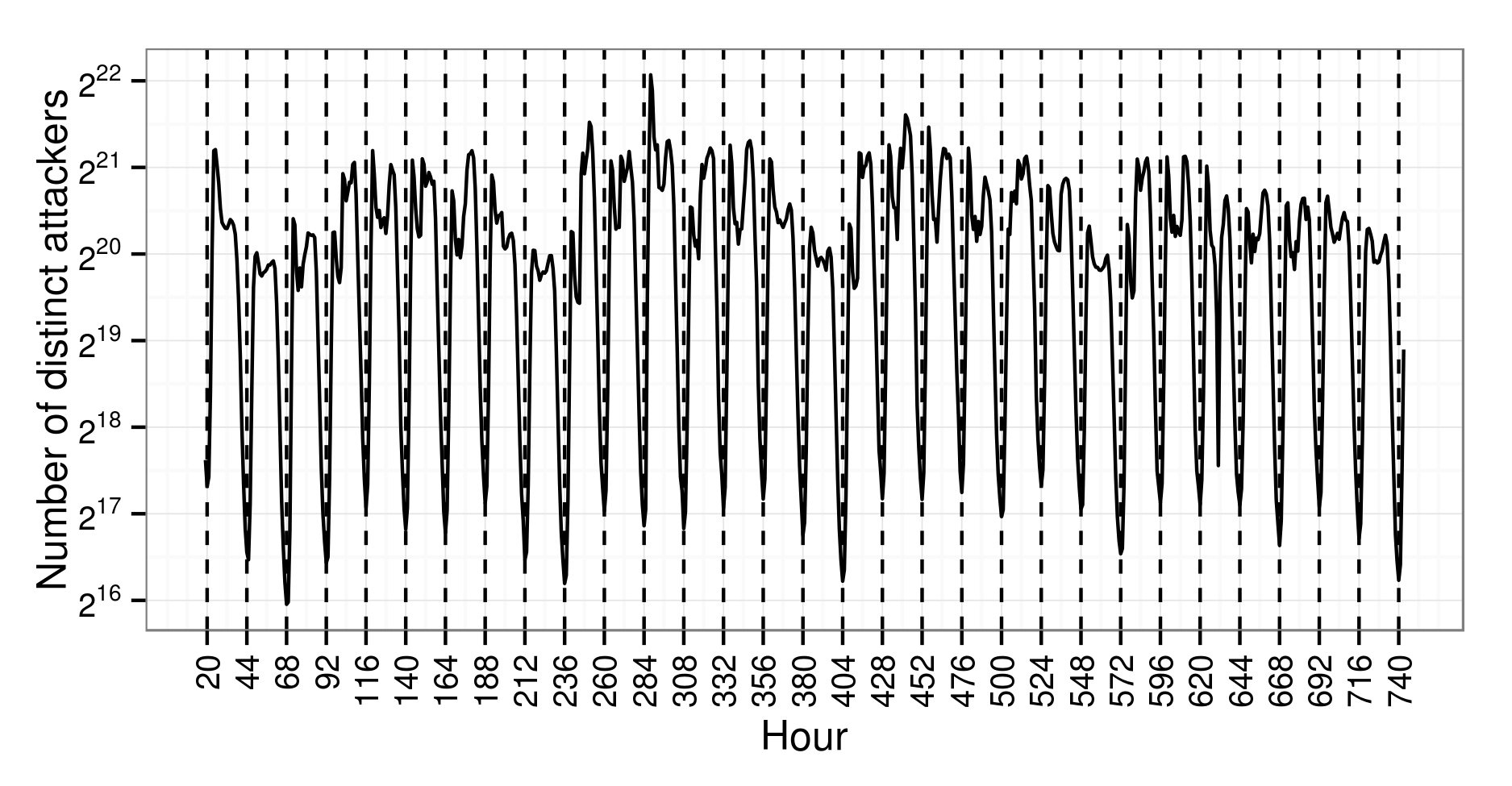}}
\caption{Elaboration of the dominance and periodicity phenomenon
(time resolution: hour).
\label{fig:vic-uniq-atker-stat}}
\end{figure}

Figure \ref{fig:vic-uniq-atker-stat-general}
further suggests a surprising consistency between the two time series. Specifically, when the number of attackers from country $X$ is large (small),
the total number of attackers is large (small).
For $D_1$, this is confirmed by Figures \ref{fig:ts-uniqAtker-waveBase} and \ref{fig:ts-uniqAtker-waveBase-CN},
which clearly show that the same periodicity is exhibited by the total number of attackers and by the number of attackers from country $X$.
For $D_2$, this is confirmed by Figures \ref{fig:D2-ts-uniqAtker-waveBase} and \ref{fig:D2-ts-uniqAtker-waveBase-CN},
which clearly show that the same periodicity is exhibited by the total number of attackers and by the number of attackers from country $X$.
We observe that the wave bases are periodic with a period of 24 hours.
After looking into the time zone of country $X$,
we find that the wave bases (i.e., that least number of attackers) correspond to the hour between 12:00 noon and 1 pm local time.
One may speculate that this is caused by computers possibly being put into the hibernate mode (during lunch time).
This may not be true because during the night hours, more computers would be put into the hibernate mode (or even powered off) and therefore even fewer attackers would be observed.
However, this is not shown by the data. One perhaps more plausible explanation is that the attacking computers may be coordinated or controlled
(for example) by botnets.

While we defer the detailed characterization of the phenomenon to Appendix \ref{appendix-characterization-the-attacker-dominance-phenomenon},
we summarize the phenomenon as:
\begin{phenomenon}
\label{phnomenon:dominance}
\emph{(The dominance and periodicity phenomenon exhibited by the number of attackers)}
The time series of the total number of attackers and the time series of the number of attackers from a particular country $X$
exhibit the same periodicity.
Moreover, the total number of attackers is dominated by the number of attackers from country $X$.
\end{phenomenon}

\section{Inferring Global Cyber Security Posture from Smaller Monitors}
\label{sec:inference}

In this section, we explore whether it is possible to use small network telescopes to approximate bigger telescopes,
from the perspectives of estimating/inferring the number of victims, attackers and attacks.
Answering this question is interesting on its own, and could lead to more cost-effective operations of network telescopes.

\smallskip

\noindent{\bf Methodology.}
We divide the /8 network telescope into $B$ equal-size blocks of IP addresses, where each block is called a {\em small telescope}. 
We want to know whether we can infer the number of victims (or attackers, or attacks) that are observed by the /8 telescope during time interval
$[t,t+1)$ at time resolution $H$ (i.e., per hour), denoted by $Y(H;t,t+1)$,
from the number of victims (or attackers, or attacks) observed by $b$ small telescopes, where $b<<B$, during the same time interval, denoted by $Y_1(H;t,t+1),\ldots,Y_b(H;t,t+1)$.
In other words, we want to know whether the following equation would hold:
      $${Y}(H;t,t+1)=c+\sum_{i=1}^b \phi_i Y_{i}(H;t,t+1),$$
where $c$ is some constant and $\phi_i$'s are coefficients.
Naturally, we can use the same PMAD measure to evaluate the estimation/inference error.

Whenever feasible, we want to consider all possible combinations of $b$ small telescopes. For $B=16$, there are ${16 \choose b}$ combinations;
for $B=256$, there are ${256\choose b}$ combinations.
For $B= 256$ and $b\ge 4$, the number of combinations becomes prohibitive.
This suggests that we first cluster the $B=256$ blocks into $b$ groups based on the DTW measure,
and then sample one block from each of the $b$ groups.
In a sense, this corresponds to the {\em best-case} scenario sampling because we need the prior information about the groups or clusters.
If the sample statistics cannot approximate the statistics derived from the data collected by the /8 telescope,
we can conclude that small telescopes are not as useful as the large telescope.

\begin{table}[!htbp]
\center
\begin{tabular}{|c|c|c|c|c|c||c|c|c|c|c|c|}
\hline
  & Min & Mean & Median & Max &SD                     &    & Min & Mean & Median & Max &SD \\\hline
 \multicolumn{6}{|c||}{$D_1$ with $B=16$: PMAD values} &  \multicolumn{6}{|c|}{$D_1$ with $B=256$: PMAD values}\\\hline
 $b=1$ & .0372 & .0472 & .0464 & .0682 & .0083        &   $b=1$ & .0563 & .0689 & .0678 & .1131 & .0074  \\\hline
 $b=2$ & .0230 & .0322 & .0302 & .0586 & .0067        &   $b=2$ & .0414 & .0535 & .0521 & .1129 & .0057 \\\hline
 $b=3$ & .0167 & .0232 & .0239 & .0301 & .0033        &   $b=3$ & .0330 & .0447 & .0436 & .0982 & .0049 \\\hline \hline
 \multicolumn{6}{|c||}{$D_2$ with $B=16$: PMAD values} &  \multicolumn{6}{|c|}{$D_2$ with $B=256$: PMAD values}\\\hline
$b=1$  & .0384 & .0484 & .0478 & .0702 & .0085        &   $b=1$  & .0799 & .1136 & .1155 & .1155 & .0071 \\\hline
$b=2$  & .0235 & .0340 & .0311 & .0702 & .0079        &   $b=2$  & .0731 & .1119 & .1155 & .1155 & .0096\\\hline
$b=3$  & .0167 & .0267 & .0251 & .0702 & .0066        &   $b=3$  & .0755 & .0787 & .0787 & .0839 & .0021\\\hline
\end{tabular}
\caption{PMAD-based measurement of the inference error when using $b$ (out of the $B$) small telescopes to approximate the
number of victims that are observed by the larger /8 telescope, where ``SD'' stands for standard deviation.
\label{tbl:pmad-sum-vic}}
\end{table}

\noindent{\bf Characterizing Inference Errors of Small Telescopes.}
From the perspective of inferring the number of victims from small telescopes,
Table \ref{tbl:pmad-sum-vic} summarizes the inference errors, in terms of the $\min$, mean, median and $\max$ PMAD values of
all the considered combinations of sample blocks, as well as the standard deviation of the PMAD values.
For $D_1$ and $B=16$, we observe that a single small telescope (out of the 16 telescopes of size $2^{20}$ IP addresses) would give good approximation of the
number of victims that would be obtained based on the larger /8 network telescope.
This is because the maximum PMAD errors is $.0682$, namely $6.82\%$ approximation error.
For $D_1$ and $B=256$, the mean approximation error is $6.89\%$ for $b=1$ (i.e., using one small telescope),
and $5.35\%$ for $b=2$ (i.e., using two small telescopes) and $4.47\%$ for $b=3$ (using three small telescopes).
For $D_2$ and $B=16$, we observe a similar phenomenon as in the case of $D_1$ and $B=16$.
However, for $D_2$ and $B=256$, the mean approximation errors is significantly larger than in the case of $D_1$ and $B=256$,
namely $11.36\%$, $11.19\%$ and $7.87\%$ for $b=1,2,3$, respectively.
Therefore, we can conclude that from the perspective of the number of victims,
a single telescope of size $2^{20}$ IP addresses would give approximately the same result as the telescope of size $2^{24}$,
and 3 randomly selected small telescope of size $2^{16}$ would give approximately the same result as the telescope of size $2^{24}$.
That is, the small telescope could be used instead.

Due to space limitation, we defer to Appendix \ref{appendix-inferrence} the characterizations on inferring the number of attackers from small telescopes
and on inferring the number of attacks from small telescopes.
Based on these characterizations, we draw the following insight:
\begin{insight}
For estimating the number of victims, substantially small telescopes could be used instead.
However, for estimating the number of attackers and the number of attacks, substantially small telescopes might not be sufficient.
\end{insight}

The above discrepancy between the number of victims and the numbers of attackers/attacks is possibly caused by the following:
The victims are somewhat ``uniformly'' attacked, but the attackers and attacks are far from ``uniformly'' distributed.
Moreover, a {\em single} attacker that scans the large telescope's IP address space will make it easy to estimate the number of victims from small telescopes.

\section{Limitations of the Study}
\label{sec:limitations}

The present study has several limitations, which are inherent to the data but not to the methodology we use.
First, our analysis treats each remote IP address as a unique attacker.
This is not accurate when the remote networks using Network Address Translation (NAT),
because remote attackers from the same network can be ``aggregated'' into a single attacker.
If many networks in country $X$ indeed use NAT, then the actual number of attackers from country $X$ is indeed larger, 
although the number of attacks from country $X$ is not affected by NAT.

Second, the characteristics presented in the paper inherently depend on the nature of network telescope.
For example, $D_1$ and $D_2$ still may contain some misconfiguration-caused, non-malicious traffic. Due to the lack of interactions between network telescope and remote computers
(an inherent limitation of network telescopes), it is hard to know the ground truth \cite{Gringoli:2009:GPU:1629607.1629610}.
Therefore, better filtering methods are needed so as to make the data approximate the ground truth as closely as possible.

Third, it is possible that some attackers are aware of the network telescope and therefore can instruct their attacks to bypass it.
As a consequence, the data  may not faithfully reflect the cybersecurity posture.

Fourth, the data collected by the network telescope does not contain rich enough information that would allow us to conduct deeper analysis,
such as analyzing the global characteristics of specific attacks.
Moreover, the data is a ``coarse'' sample of the ground-truth cybersecurity posture because
the first flow from a remote attacker may be a scan/probe activity, or a first attack attempt against a specific port.

\section{Conclusion}
\label{sec:conclusion}

We have studied the cybersecurity posture based on the data collected by CAIDA's network telescope.
We have found that the {\em sweep-time}
should be characterized as a stochastic process rather than a random variable.
We also have found that the {\em total} number of attackers (and attacks) that are observed by the network telescope is largely determined by
the number of attackers from a single country. 
There are many interesting problems for future research, such as:
How can we (more) accurately predict the time series? To what extent they are predictable?

\smallskip

\noindent{\bf Acknowledgement}. We thank CAIDA for sharing with us the data analyzed in the paper.
This work was supported in part by ARO Grant \#W911NF-13-1-0141 and NSF Grant \#1111925.

\begin{appendix}

\section{Characterization of the Dominance and Periodicity Phenomenon Exhibited by Attackers}
\label{appendix-characterization-the-attacker-dominance-phenomenon}

Now we quantify the similarity between the two time series via
Dynamic Time Warping (DTW),
fitted model, and prediction accuracy.

\smallskip

\begin{figure*}[hbpt]
\centering
\subfigure[{ Total in $D_1$ vs. total in $D_2$}\label{fig:ll-d1-d2-atker}]{\includegraphics[width=.24\textwidth]{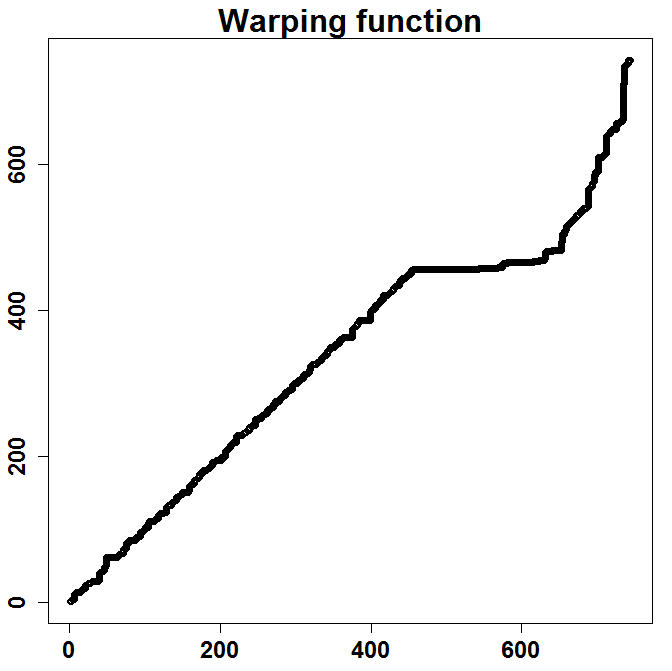}}
\subfigure[{ Total in $D_1$ vs. country $X$ in $D_1$}\label{fig:wraping}]{\includegraphics[width=.24\textwidth]{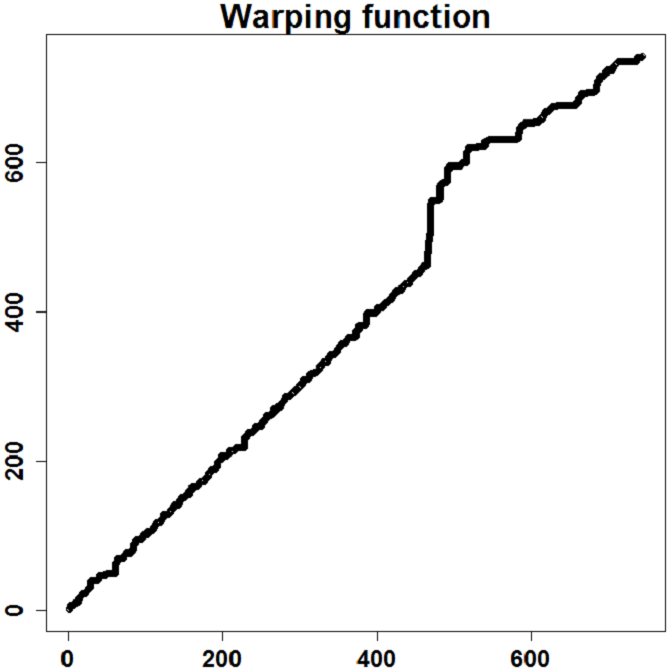}}
\subfigure[{ Country $X$ in $D_1$ vs. country $X$ in $D_2$}\label{fig:cn-d1-d2-atker}]{\includegraphics[width=.24\textwidth]{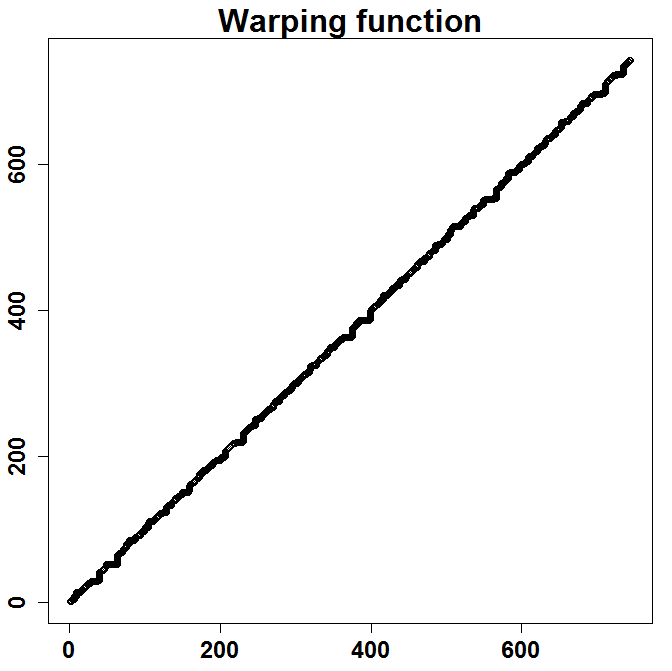}}
\subfigure[{ Total in $D_2$ vs. country $X$ in $D_2$}\label{fig:D2-wraping}]{\includegraphics[width=.24\textwidth]{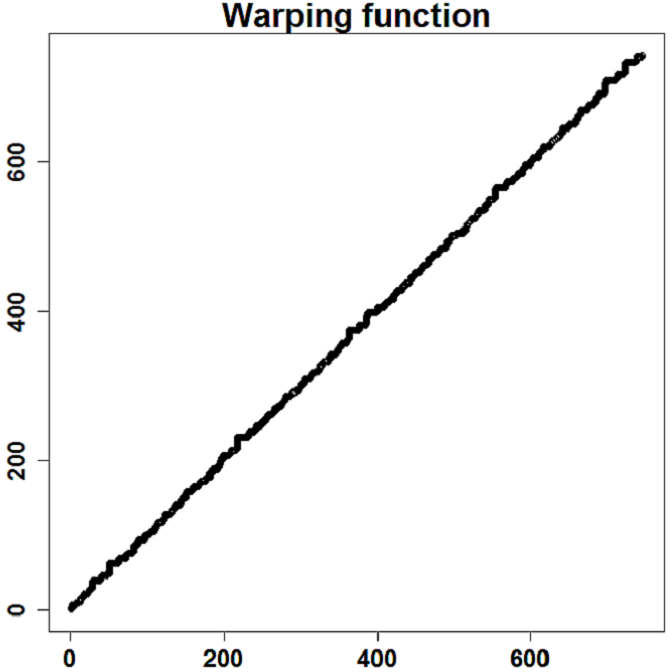}}
\caption{DTW statistics between the times series of the total number of attackers and the time series of the number of attackers for country $X$.
\label{fig:dtw-s-D1-D2}}
\end{figure*}

\noindent{\bf Similarity based on DTW.}
Figure \ref{fig:ll-d1-d2-atker} plots the warping path between the total number attackers in $D_1$ and the total number of  attackers in $D_2$.
The two time series are very similar to each other, except for the time interval $[452,668]$
as suggested by Figures \ref{fig:ts-uniqAtker-waveBase} and \ref{fig:D2-ts-uniqAtker-waveBase}.
Figure \ref{fig:wraping} plots the warping path between the two time series plotted in Figure \ref{fig:vic-uniq-atker-cc-2},
namely the total number of attackers in $D_1$ and the total number of attackers from country $X$ in $D_1$.
It shows that the two time series are very similar to each other except during the time interval $[455, 630]$, as suggested
by Figures \ref{fig:ts-uniqAtker-waveBase} and \ref{fig:ts-uniqAtker-waveBase-CN}.
Figure \ref{fig:cn-d1-d2-atker} plots the warping path between the two time series plotted in Figures \ref{fig:ts-uniqAtker-waveBase-CN} and \ref{fig:D2-ts-uniqAtker-waveBase-CN}.
It shows that the two time series are almost identical to each other, and that the filtering of rarely seen attackers/attacks does not manipulate the periodic structure of
the time series of the number of attackers from country $X$.
Figure \ref{fig:D2-wraping} plots the warping path between the two time series plotted in Figures \ref{fig:D2-ts-uniqAtker-waveBase} and \ref{fig:D2-ts-uniqAtker-waveBase-CN},
which indeed are almost identical to each other.

\noindent{\bf Similarity based on fitted models.}
Since both the time series exhibit periodicity,
we use the multiplicative seasonal ARIMA model to fit the two time series in $D_1$ and $D_2$, respectively.
The model parameters are: nonseasonal orders $(p,d,q)$, and seasonal orders $(P,D,Q)$, and seasonal period $s=24$ based on the above discussion of periodicity.
For model selection, the parameter sets are:
\begin{itemize}
  \item  $(p,d,q)\in [0,5]\times \{0,1\}\times [0,5]$;
  \item  $(P,D,Q)\in [0,5]\times \{0,1\}\times [0,5]$.
\end{itemize}
According to the AIC criterion (briefly reviewed in Section \ref{sec:statistical-preliminaries}),
the two time series in both $D_1$ and $D_2$ prefer to the following model:
\begin{eqnarray*}
W_t&=&\phi_1 W_{t-1}+e_t+\theta_1 e_{t-1}+\Phi_1 W_{t-24}+\Phi_2 W_{t-48}+ \\
&&\Theta_1 e_{t-24}+\Theta_2 e_{t-48}+\Theta_3 e_{t-96},
\end{eqnarray*}
where $W_t=|A(r;t,t+1)|-|A(r;t-24,t-23)|$.
Table \ref{tbl:D1-D2-country-total-fitting} summarizes the fitting results.
We observe that the two fitted models in $D_1$ are similar to each other in terms of coefficients,
and that the two fitted models in $D_2$ are almost identical to each other.

\begin{table}[!htbp]
\center
{\small
\begin{tabular}{|c||r|r|r|r|r|r|r|}
\hline
 & $\phi_1$ & $\theta_1$ & $\Phi_1$  & $\Phi_2$ & $\Theta_1$ & $\Theta_2$ & $\Theta_3$ \\\hline\hline
\multicolumn{8}{|c|}{Fitted model of total number of attackers in $D_1$: PMAD=.08}\\\hline
Coefficients  & .91 &  .38  &  1.22 &  -.98 & -2.15 & 2.11 &  -.86 \\\hline\hline
\multicolumn{8}{|c|}{Fitted model of number of  attackers from country $X$ in $D_1$: PMAD=.06}\\\hline
Coefficients &  0.82 & .39 & 1.22 & -.99 & -2.19 & 2.16 & -.91\\\hline
\hline
\multicolumn{8}{|c|}{Fitted model of total number of attackers in $D_2$: PMAD=.08}\\\hline
Coefficients  & .79 & .4 &  1.21 & -.99 & -2.18 &  2.16 & -.9 \\\hline\hline
\multicolumn{8}{|c|}{Fitted model of number of  attackers from country $X$ in $D_2$: PMAD=.07}\\\hline
Coefficients & .79 & .4 & 1.21 & -.99 & -2.19 & 2.16 & -.9\\\hline
\end{tabular}
\caption{Coefficients in the fitted models of
the total number of attackers and the number of attackers from country $X$.
\label{tbl:D1-D2-country-total-fitting}}
}
\end{table}

\noindent{\bf Similarity based on prediction accuracy.}
Table \ref{tbl:country-pred} summarizes the PMAD values for 1, 4, 7 and 10 hours ahead-of-time prediction of the number of attackers during the last 96 hours in both $D_1$ and $D_2$.
For $D_1$, we observe that 1-hour ahead-of-time predictions for the number of attackers from country $X$ and the total number of attackers are reasonably accurate (with PMAD value
.093 and .092, or $9.3\%$ and $9.2\%$ prediction error, respectively); whereas the predictions for 4, 7 and 10 hours ahead-of-time are not useful.
For $D_2$, we observe similar prediction results, namely that 1-hour ahead-of-time
predictions lead to $7.5\%$ prediction error for the total number of attackers
and $9.5\%$ prediction error for the number of attackers from country $X$.

\begin{table}[!htbp]
\center
\begin{tabular}{|c|c|c|c|c||c|c|c|c|c|}
\hline
 & $h=1$ & $h=4$ & $h=7$ & $h=10$           &  & $h=1$ & $h=4$ & $h=7$ & $h=10$  \\\hline
 \multicolumn{5}{|c||}{$D_1$: PMAD values}   & \multicolumn{5}{|c|}{$D_2$: PMAD values} \\\hline
Total & .092 & .244 & .333  & .404          & Total & .075 & .156 & .180  & .177 \\\hline
Country $X$ & .093  & .208 &  .240 & .245   &  Country $X$ & .095  & .203  & .230  & .224\\\hline
\end{tabular}
\caption{PMAD values for $h=1,4,7,10$ hours ahead-of-time predictions on the total number of attackers and on the number of attackers from country $X$,
as observed by the telescope. \label{tbl:country-pred}}
\end{table}

\section{Further Characterizations on the Inference Errors of Small Telescopes}
\label{appendix-inferrence}

\noindent{\bf Inferring the number of attackers from small telescopes.}
Similarly, we would like to infer the number of attackers based on small telescopes.
Table \ref{tbl:pmad-sum-aker} summarizes the inference errors in terms of the $\min$, mean, median and $\max$ PMAD values of
all the considered combinations of sample blocks, as well as the standard deviation of the PMAD values.
For $D_1$ and $B=16$, we observe that 3 small telescopes
(out of the 16 telescopes of size $2^{20}$ IP addresses) would give good approximation of the
number of attackers that would be obtained based on the network telescope of size $2^{24}$.
This is because the maximum PMAD value is $7.34\%$.
For $D_2$ and $B=16$, we observe that using 4 small telescopes of size $2^{20}$ does not lead to good approximation.
For $B=256$, neither $D_1$ nor $D_2$ leads to obtain good enough approximation.
These suggest that using significantly small telescopes may not lead to robust results.

\begin{table}[!htbp]
\center
\begin{tabular}{|c|c|c|c|c|c||c|c|c|c|c|c|}
\hline
  & Min & Mean & Median & Max &SD                             &    & Min & Mean & Median & Max &SD \\\hline
 \multicolumn{6}{|c||}{$D_1$ with $B=16$: PMAD values}        &  \multicolumn{6}{|c|}{$D_2$ with $B=16$: PMAD values}\\\hline
 $b=1$  & .0663 & .0921 & .0916 & .1329 & .0205               &  $b=1$ & .1689 & .1863 & .1857 & .2221 & .0119 \\\hline
 $b=2$  & .0626 & .0797 & .0752 & .1194 & .0138               &  $b=2$ & .1476 & .1784 & .1798 & .2221 & .0089\\\hline
 $b=3$  & .0624 & .0700 & .0713 & .0734 & .0033               &  $b=3$ & .1395 & .1730 & .1755 & .2221 & .0098\\\hline
 $b=4$  & .0593 & .0685 & .0693 & .0734 & .0036               &  $b=4$ & .1355 & .1664 & .1676 & .1874 & .0103\\\hline
 \multicolumn{6}{|c||}{$D_1$ with $B=256$: PMAD values}        &  \multicolumn{6}{|c|}{$D_2$ with $B=256$: PMAD values}\\\hline
 $b=1$ & .1273 & .2499 & .3037 & .4303 & .0983                &  $b=1$ & .2967 & .4346 & .4387 & .4396 & .0172\\\hline
 $b=2$ & .1121 & .1929 & .1467 & .4303 & .0846                &  $b=2$ & .1712 & .4307 & .4387 & .4396 & .0240\\\hline
 $b=3$ & .1251 & .1510 & .1447 & .3287 & .0232                &  $b=3$ & .1713 & .2369 & .2352 & .3730 & .0422\\\hline
 $b=4$ & .1250 & .1491 & .1445 & .2935 & .0188                &  $b=4$ & .1664 & .2261 & .2329 & .2542 & .0231\\\hline
 $b=5$ & .1236 & .1474 & .1435 & .2762 & .0168                &  $b=5$ & .1618 & .1658 & .1664 & .1674 & .0015\\\hline
\end{tabular}
\caption{PMAD-based measurement of the inference error when using $b$ (our of the $B$) small telescopes to approximate the
number of attackers that are observed by the larger /8 telescope, where ``SD'' stands for standard deviation.
\label{tbl:pmad-sum-aker}}
\end{table}

\noindent{\bf Inferring the number of attacks from small telescopes.}
From the perspective of inferring the number of attacks,
Table \ref{tbl:pmad-sum-att} summarizes the inference errors as in the above.
For $D_1$ and $B=16$, we observe that 3 small telescopes
(out of the 16 telescopes of size $2^{20}$ IP addresses) would give good approximation of the
number of attacks that would be obtained based on the larger network telescope of size $2^{24}$.
This is because the maximum PMAD errors is $.0761$, namely $7.61\%$ approximation error.
For $D_1$ and $B=256$, the mean approximation error is $9.58\%$ for $b=5$ (i.e., using 5 small telescopes instead), which is marginally acceptable.
For $D_2$ and $B=16$, we observe that using 4 small telescopes of size $2^{20}$ can lead to worst-case approximation error $8.27\%$.
For $D_2$ and $B=256$, we observe that using 5 small telescopes of size $2^{16}$ does not lead to good approximation.
That is, substantially small telescope may not be as useful as the large telescope.

\begin{table}[!htbp]
\center
\begin{tabular}{|c|c|c|c|c|c||c|c|c|c|c|c|}
\hline
  & Min & Mean & Median & Max &SD                               &   & Min & Mean & Median & Max &SD \\\hline
 \multicolumn{6}{|c||}{$D_1$ with $B=16$: PMAD values}          &  \multicolumn{6}{|c|}{$D_2$ with $B=16$: PMAD values}\\\hline
 $b=1$  & .0755  & .0890 & .0883 & .1106 & .0095                &  $b=1$ & .0765 & .0901 & .0895 & .1130 & .0096   \\\hline
 $b=2$  & .0436  & .0629 & .0589 & .0953 & .0138                &  $b=2$ & .0444 & .0655 & .0659 & .1130 & .0150 \\\hline
 $b=3$  & .0363  & .0488 & .0435 & .0761 & .0125                &  $b=3$ & .0343 & .0518 & .0461 & .1130 & .0138  \\\hline
 $b=4$  & .0272  & .0392 & .0369 & .0739 & .0100                &  $b=4$ & .0267 & .0395 & .0372 & .0827 & .0094\\\hline
 \multicolumn{6}{|c||}{$D_1$ with $B=256$: PMAD values}         &  \multicolumn{6}{|c|}{$D_2$ with $B=256$: PMAD values}\\\hline
 $b=1$ & .1166 & .1322 & .1311 & .1917 & .0098                  &  $b=1$ & .1344 & .1909 & .1938 & .1938 & .0117\\\hline
 $b=2$ & .0853 & .1069 & .1056 & .1917 & .0092                  &  $b=2$ & .1197 & .1881 & .1938 & .1938 & .0160 \\\hline
 $b=3$ & .0699 & .0916 & .0897 & .1917 & .0086                  &  $b=3$ & .1323 & .1409 & .1387 & .1602 & .0079\\\hline
 $b=4$ & .0835 & .0984 & .0970 & .1700 & .0069                  &  $b=4$ & .1310 & .1383 & .1366 & .1520 & .0069\\\hline
 $b=5$ & .0818 & .0958 & .0945 & .1582 & .0064                  &  $b=5$ & .1256 & .1312 & .1310 & .1419 & .0048\\\hline
\end{tabular}
\caption{PMAD-based measurement of the inference error when using $b$ (our of the $B$) small telescopes to approximate the
number of attacks observed by the /8 telescope, where ``SD'' stands for standard deviation.
\label{tbl:pmad-sum-att}}
\end{table}

\end{appendix}

\end{document}